\documentclass[preprintnumbers,unsortedaddress,superscriptaddress,notitlepage]{revtex4}
\usepackage{amsfonts}
\usepackage{amsmath}
\usepackage{hyperref}
\usepackage{amssymb}
\usepackage[english]{babel}
\usepackage{graphicx}
\usepackage{epsfig}
\usepackage{bm}
\usepackage{longtable}
\usepackage{verbatim}
\usepackage{longtable}
\usepackage[usenames,dvipsnames]{color}
\usepackage{multirow}

\newcommand{\mathsym}[1]{{}}

\newcommand{\emp}{\begin{equation}}
\newcommand{\fin}{\end{equation}}
\newcommand{\empn}{\begin{equation*}}
\newcommand{\finn}{\end{equation*}}

\newcommand{\bea}{\begin{eqnarray}}
\newcommand{\eea}{\end{eqnarray}}
\newcommand{\eger}{\begin{gather}}
\newcommand{\fger}{\end{gather}}
\newcommand{\egn}{\begin{gather*}}
\newcommand{\fgn}{\end{gather*}}
\newcommand{\bit}{\begin{itemize}}
\newcommand{\eit}{\end{itemize}}

\newcommand{\R}{\mca R}

\input{tcilatex}
\begin{document}

\title{Adjoint $SU(5)$ GUT model with $T_{7}$ flavor symmetry}
\author{Carolina Arbel\'aez}
\email{carbela.arbelaez@usm.cl}
\affiliation{{\small Universidad T\'ecnica Federico Santa Mar\'{\i}a and Centro Cient%
\'{\i}fico-Tecnol\'ogico de Valpara\'{\i}so}\\
Casilla 110-V, Valpara\'{\i}so, Chile}
\author{A. E. C\'arcamo Hern\'andez}
\email{antonio.carcamo@usm.cl}
\affiliation{{\small Universidad T\'ecnica Federico Santa Mar\'{\i}a and Centro Cient%
\'{\i}fico-Tecnol\'ogico de Valpara\'{\i}so}\\
Casilla 110-V, Valpara\'{\i}so, Chile}
\author{Sergey Kovalenko}
\email{sergey.kovalenko@usm.cl}
\affiliation{{\small Universidad T\'ecnica Federico Santa Mar\'{\i}a and Centro Cient%
\'{\i}fico-Tecnol\'ogico de Valpara\'{\i}so}\\
Casilla 110-V, Valpara\'{\i}so, Chile}
\date{\today }
\author{Iv\'an Schmidt}
\email{ivan.schmidt@usm.cl}
\affiliation{{\small Universidad T\'ecnica Federico Santa Mar\'{\i}a and Centro Cient%
\'{\i}fico-Tecnol\'ogico de Valpara\'{\i}so}\\
Casilla 110-V, Valpara\'{\i}so, Chile}

\begin{abstract}
We propose an adjoint $SU(5)$ GUT model 
with a $T_{7}$ family symmetry and 
an extra $Z_{2}\otimes Z_{3}\otimes Z_{4}\otimes Z_{4}^{\prime }\otimes
Z_{12}$ discrete group, that successfully describes the prevailing Standard
Model fermion mass and mixing pattern. The observed hierarchy of the charged
fermion masses and the quark mixing angles arises from the $Z_{3}\otimes
Z_{4}\otimes Z_{12}$ symmetry breaking, which occurs near 
the GUT scale. The light active neutrino masses are generated by type-I and
type-III seesaw mechanisms mediated by the fermionic $SU(5)$ singlet and 
the adjoint $\mathbf{24}$-plet. 
The model predicts 
the effective Majorana neutrino mass parameter of 
neutrinoless double beta decay 
to be $m_{\beta \beta }=$ 4 and 50 meV for the normal and the inverted
neutrino spectra, respectively. 
We construct several benchmark scenarios, 
which 
lead to $SU(5)$ gauge coupling unification and 
are compatible with the known phenomenological constraints originating from
the lightness of neutrinos, 
proton decay, dark matter 
etc. 
These 
scenarios contain TEV-scale colored fields, 
which could 
give rise 
to a visible signal or be stringently constrained at the LHC. 
\end{abstract}

\maketitle


\def\2tvec#1#2{
\left(
\begin{array}{c}
#1  \\
#2  \\   
\end{array}
\right)}

\def\mat2#1#2#3#4{
\left(
\begin{array}{cc}
#1 & #2 \\
#3 & #4 \\
\end{array}
\right)
}

\def\Mat3#1#2#3#4#5#6#7#8#9{
\left(
\begin{array}{ccc}
#1 & #2 & #3 \\
#4 & #5 & #6 \\
#7 & #8 & #9 \\
\end{array}
\right)
}

\def\3tvec#1#2#3{
\left(
\begin{array}{c}
#1  \\
#2  \\   
#3  \\
\end{array}
\right)}

\def\4tvec#1#2#3#4{
\left(
\begin{array}{c}
#1  \\
#2  \\   
#3  \\
#4  \\
\end{array}
\right)}

\def\5tvec#1#2#3#4#5{
\left(
\begin{array}{c}
#1  \\
#2  \\
#3  \\
#4  \\
#5  \\
\end{array}
\right)}

\def\L{\left}
\def\R{\right}

\def\pl{\partial}

\def\lra{\leftrightarrow}


\section{Introduction}

The LHC discovery of a $126$ GeV Higgs boson \cite{LHC-H-discovery} has
completed the era of the experimental quest for the missing elements of the
Standard Model (SM). Now it is a well- established and extremely successful
theory of the electroweak phenomena. 
However, the SM has several unaddressed issues, such as 
the smallness of neutrino masses, 
the observed pattern of fermion masses and mixings, and the existence of
three fermion families 
\cite{PDG}, etc. 
The observed pattern of fermion masses spans over a range of 5 orders of
magnitude in the quark sector and 
much wider considering neutrinos. 
Neutrino oscillation experiments 
demonstrated that at least two of the neutrinos are massive, with masses
much lower, by 
several orders of magnitude, 
than the other SM 
fermions, and also that all of the three neutrino flavors mix with each
other. 
The smallness of quark mixing 
contrasts with the sizable 
mixing of neutrinos; 
i.e., while in the quark sector all the mixing angles are small, in the 
neutrino sector two of them 
are large, and only one mixing angle is small. This suggests that the
neutrino sector is described by a different kind of underlying physics than
the quark sector. As is well known the tiny neutrino masses may point
towards a high-energy scale of New Physics, where lepton number violation
(LNV) takes place.


The physical observables in the neutrino sector, i.e., the neutrino
mass-squared splittings and mixing parameters, are constrained from the
global fits of the available data from neutrino oscillation experiments Daya
Bay \cite{An:2012eh}, T2K \cite{Abe:2011sj}, MINOS \cite{Adamson:2011qu},
Double CHOOZ \cite{Abe:2011fz}, and RENO \cite{Ahn:2012nd}, as shown in
Tables \ref{NH} and \ref{IH} (based on Ref. \cite{Forero:2014bxa}) for the
normal (NH) and inverted (IH) hierarchies of the neutrino mass spectrum. As
seen from these tables %
the neutrino oscillation experimental data show a clear evidence of a
violation of the so-called tribimaximal symmetry described by the
tribimaximal mixing matrix (TBM), 
predicting the neutrino mixing angles 
$\left( \sin ^{2}\theta _{12}\right) _{TBM}=\frac{1}{3}$, $\left( \sin
^{2}\theta _{23}\right) _{TBM}=\frac{1}{2}$, and $\left( \sin ^{2}\theta
_{13}\right) _{TBM}=0$. 
\begin{table}[tbh]
\begin{tabular}{|c|c|c|c|c|c|}
\hline
Parameter & $\Delta m_{21}^{2}$($10^{-5}$eV$^2$) & $\Delta m_{31}^{2}$($%
10^{-3}$eV$^2$) & $\left( \sin ^{2}\theta _{12}\right) _{\exp }$ & $\left(
\sin ^{2}\theta _{23}\right) _{\exp }$ & $\left( \sin ^{2}\theta
_{13}\right) _{\exp }$ \\ \hline
Best fit & $7.60$ & $2.48$ & $0.323$ & $0.567$ & $0.0234$ \\ \hline
$1\sigma $ range & $7.42-7.79$ & $2.41-2.53$ & $0.307-0.339$ & $0.439-0.599$
& $0.0214-0.0254$ \\ \hline
$2\sigma $ range & $7.26-7.99$ & $2.35-2.59$ & $0.292-0.357$ & $0.413-0.623$
& $0.0195-0.0274$ \\ \hline
$3\sigma $ range & $7.11-8.11$ & $2.30-2.65$ & $0.278-0.375$ & $0.392-0.643$
& $0.0183-0.0297$ \\ \hline
\end{tabular}%
\caption{Range for experimental values of neutrino mass squared splittings
and leptonic mixing parameters, taken from Ref. \protect\cite{Forero:2014bxa}%
, for the case of normal hierarchy.}
\label{NH}
\end{table}
\begin{table}[tbh]
\begin{tabular}{|c|c|c|c|c|c|}
\hline
Parameter & $\Delta m_{21}^{2}$($10^{-5}$eV$^{2}$) & $\Delta m_{13}^{2}$($%
10^{-3}$eV$^{2}$) & $\left( \sin ^{2}\theta _{12}\right) _{\exp }$ & $\left(
\sin ^{2}\theta _{23}\right) _{\exp }$ & $\left( \sin ^{2}\theta
_{13}\right) _{\exp }$ \\ \hline
Best fit & $7.60$ & $2.38$ & $0.323$ & $0.573$ & $0.0240$ \\ \hline
$1\sigma $ range & $7.42-7.79$ & $2.32-2.43$ & $0.307-0.339$ & $0.530-0.598$
& $0.0221-0.0259$ \\ \hline
$2\sigma $ range & $7.26-7.99$ & $2.26-2.48$ & $0.292-0.357$ & $0.432-0.621$
& $0.0202-0.0278$ \\ \hline
$3\sigma $ range & $7.11-8.11$ & $2.20-2.54$ & $0.278-0.375$ & $0.403-0.640$
& $0.0183-0.0297$ \\ \hline
\end{tabular}%
\caption{Range for experimental values of neutrino mass squared splittings
and leptonic mixing parameters, taken from Ref. \protect\cite{Forero:2014bxa}%
, for the case of inverted hierarchy.}
\label{IH}
\end{table}

Addressing the flavor puzzle 
requires 
extensions of the SM, including larger scalar and/or fermion sectors, as
well as an extended gauge group with additional flavor symmetries, which
allow one to explain the SM fermion mass and mixing pattern. Along this
line, several models have been proposed in the literature (for a review see,
e.g., Refs. \cite%
{Fritzsch:1999ee,Altarelli:2002hx,Altarelli:2010gt,Ishimori:2010au,FlavorSymmRev}%
). 
The fermion mass and mixing pattern can also be described by postulating
particular mass matrix textures (see, e.g., Ref. \cite{Textures} for a
comprehensive review and some recent works considering textures). 
It is believed that grand unified theories (GUTs) endowed with flavor
symmetries could provide an unified description for the mass and mixing
pattern of leptons and quarks. 
This is motivated by the fact that leptons and quarks belong to the same
multiplets of the GUT group, allowing to relate their masses and mixings 
\cite{Marzocca:2011dh,Antusch:2013kna,Fichet:2014vha}. Furthermore, this
setup can generate small neutrino masses through a type-I seesaw mechanism,
where the new heavy Majorana neutrinos acquire very large masses due to
their interactions with scalar singlets, assumed to get vacuum expectation
values (VEVs) at the very high energy scale. 
Various GUT models with flavor symmetries have been proposed in the
literature 
\cite%
{Chen:2013wba,King:2012in,Meloni:2011fx,BhupalDev:2012nm,Babu:2009nn,Babu:2011mv,Gomez-Izquierdo:2013uaa,Antusch:2010es,Hagedorn:2010th,Ishimori:2008fi,Patel:2010hr,Cooper:2010ik,Emmanuel-Costa:2013gia,Chen:2007,Antusch:2014poa,Gehrlein:2014wda,Campos:2014lla,Carcamo:2014,Gehrlein:2015dxa,Bjorkeroth:2015ora}%
. 
For a general review, see for example \cite{King:2013eh,Chen:2003zv}.

\quad In this paper we propose a version of the adjoint $SU(5)$ GUT model,
where an extra 
\mbox{$Z_{2}\otimes Z_{3}\otimes Z_{4}\otimes Z_{4}^{\prime }\otimes
Z_{12}$} discrete group (see Ref \cite{T7} for a comprehensive study of 
the $T_{7}$ flavor group) extends the symmetry of the model, and several
scalar fields are included to generate viable and predictive textures for
the fermion sector. 
The particular role of each additional scalar field and the corresponding
particle assignments under the symmetry group of the model are explained in
detail in Sec. \ref{model}. The fermionic sector of our model, in addition
to the SM fermions, 
contains one heavy Majorana neutrino $N_{R}$, singlet under the SM group,
and an adjoint $\mathbf{24}$ fermionic irreducible representation (irrep) of 
$SU(5)$ 
so that the light neutrino masses are generated via type-I and type-III
seesaw mechanisms. 
An adjoint $SU(5)$ GUT model, but without discrete symmetries, and more
minimal particle content 
was previously considered in Ref. \cite{Perez:2007rm}. 
%
Although our model is less minimal than that of Ref.\cite{Perez:2007rm}, it
provides a successful description of the SM fermion mass and mixing pattern,
not addressed in Ref.\cite{Perez:2007rm}. Also, our model is more predictive
in the leptonic sector. 
The model has 14 free effective parameters, which allow us 
to reproduce the experimental values of 18 observables with good accuracy:
i.e., 9 charged fermion masses, 2 neutrino mass-squared splittings, 3 lepton
mixing parameters, and 4 parameters of the Wolfenstein CKM quark mixing
matrix parametrization. 
It is noteworthy that the alternative $SU(5)$ GUT model of Ref. \cite%
{Antusch:2010es}, with a supersymmetric setup and flavor symmetries, also
has 14 free effective parameters aimed at reproducing the above mentioned 18
observables. Discrete symmetries different from $T_{7}$ have also been
employed in adjoint $SU(5)$ GUT models. Some examples are the supersymmetric
adjoint $SU(5)$ GUT model with $A_{4}$ flavor symmetry of Ref. \cite%
{Cooper:2010ik}, and the nonsupersymmetric $SU(5)$ GUT model with $Z_{4}$
symmetry of Ref. \cite{Emmanuel-Costa:2013gia}. The aforementioned SUSY
adjoint $SU(5)$ GUT model with $A_{4}$ flavor symmetry employs two sets of $%
Z_{2}$ symmetries and two sets of $U(1)$ symmetries. One of these $U(1)$
symmetries is global and represents an $R$-parity symmetry, whereas the
other $U(1)$ symmetry is assumed to be gauged. The two $Z_{2}$ symmetries
shape the Yukawa matrices for quarks and charged leptons. That model
includes Higgs multiplets in $\mathbf{1}$, $\mathbf{5}$, $\overline{\mathbf{5%
}}$, $\overline{\mathbf{45}}$, dimensional representations of $SU(5)$. The 14 
$SU(5)$ scalar singlets are grouped into three $A_{4}$ triplets and two $%
A_{4}$ trivial singlets. As in our model, neutrino masses arise from a
combination of type-I and type-III seesaw mechanisms. Whereas in the model
of Ref. \cite{Cooper:2010ik}, the CKM quark mixing matrix mainly arises from
the down-type quark sector, in our model, the quark mixing is completely
determined from the up-type quark sector. Furthermore, in this model the
leptonic mixing angles are determined by two parameters, whereas in our
model only one parameter determines the leptonic mixing angles. Moreover,
the quark masses and mixings are studied in detail in our model, whereas in
the model of Ref. \cite{Cooper:2010ik} a detailed study of quark masses and
mixings is not included. Besides that, our model includes a detailed
discussion of gauge coupling unification and seesaw mass scale limits, not
performed in the model of Ref. \cite{Cooper:2010ik}. With respect to the
nonsupersymmetric adjoint $SU(5)$ GUT model with $Z_{4}$ discrete symmetry
of Ref. \cite{Emmanuel-Costa:2013gia}, the discrete symmetry is introduced
in that model in order to generate the nearest-neighbour-interaction
textures for charged fermions. The scalar sector of that model includes an
adjoint multiplet, a quintuplet and one $\mathbf{45}$-dimensional
representation, and neutrino masses arise from type-I seesaw, type-III
seesaw, and one-loop radiative seesaw mechanisms. Despite the fact that this
model is more minimal than our $T_{7}$ flavor adjoint $SU(5)$ GUT model, our
model has a much more predictive lepton sector. Let us note that the lepton
sector of the model of Ref. \cite{Emmanuel-Costa:2013gia} has 12 effective
free parameters, whereas the lepton sector of our model has a total of 6
effective free parameters. Unlike the models of Refs. \cite%
{Cooper:2010ik,Emmanuel-Costa:2013gia}, our adjoint $SU(5)$ GUT model, which
is non supersymmetric, employs a {$Z_{2}\otimes Z_{3}\otimes Z_{4}\otimes
Z_{4}^{\prime }\otimes Z_{12}$} discrete symmetry.

\quad After imposing gauge coupling unification 
equal to or better than that in the MSSM (another difference with respect to
Refs. \cite{Perez:2007rm,FileviezPerez:2007nh}, which impose exact gauge
coupling unification), we find a wide set of simple field configurations,
which pass 
proton decay constraints and give rise to the neutrino masses through a
type-I and type-III seesaw realization. Considering the limit on the triplet
scalar mass (denoted here as $\Xi _{3}$), which comes from the cold dark
matter constraints pointed out in Ref. \cite{Perez:2008ry}, we found lower
limits on the seesaw scale for two simple scenarios with different sets of
beyond-the-SM field 
configurations. 
%

\quad The paper is organized as follows. In Sec. \ref{model} we outline the
proposed model. In Sec. \ref{massmix} we present our results on 
neutrino masses and mixing, 
followed by a numerical analysis. Our results for the quark sector, with the
corresponding numerical analysis, are presented in Sec. \ref{Quarkmixing}.
Gauge coupling unification and seesaw scale mass limits are discussed in Sec %
\ref{Unification}. We conclude with discussions and a summary in Sec. \ref%
{Summary}. Some necessary facts about the $T_{7}$ group and details of our
analysis are collected in appendixes.

\section{The Model\label{model}}

\label{model} 
The first grand unified theory (GUT) proposed in Ref. \cite{Georgi:1974sy} 
is based on the $SU(5)$ gauge symmetry accommodating the SM 
fermions in $\mathbf{\bar{5}}+\mathbf{10}$ and the scalars in $\mathbf{5}+%
\mathbf{24}$ irreps of $SU(5)$. As is well known this model suffers from
several problems. In particular, it predicts wrong down-type quark and
charged lepton mass relations and a short proton lifetime, and the
unification of gauge couplings disagrees with the values of $\alpha _{S}$, $%
\sin \theta _{W}$ and $\alpha _{em}$ measured at the $M_{Z}$ scale. Moreover
this minimal $SU(5)$ GUT model does not include a mechanism for generating
nonvanishing neutrino masses in contradiction with experimental data on
neutrino oscillations. 
The minimal $SU(5)$ GUT model can be improved 
by including, in particular, a scalar $\mathbf{45}$ irrep of $SU(5)$ \cite%
{Georgi:1979df,Frampton:1979,Ellis:1979,Nandi:1980sd,Frampton:1980,Langacker:1980js,Kalyniak:1982pt,Giveon:1991,Dorsner:2007fy,Dorsner:2006dj,FileviezPerez:2007nh,Perez:2008ry,Khalil:2013ixa}%
. 
This next-to-minimal $SU(5)$ GUT model fails, however, in describing the
observed pattern of fermion masses and mixings, due to the lack of
explanation for the hierarchy among the large number of Yukawa couplings in
the model. Below 
we consider a multi-Higgs extension of the next-to-minimal $SU(5)$ GUT
model, which successfully describes the 
pattern of the SM fermion masses and mixing. The 
full symmetry $\mathcal{G}$ of the model is broken in two subsequent steps
as follows: 
\begin{eqnarray}  \label{Symmetry-1}
&&\mathcal{G}=SU\left( 5\right) \otimes T_{7}\otimes Z_{2}\otimes
Z_{3}\otimes Z_{4}\otimes Z_{4}^{\prime }\otimes Z_{12}  \label{Group} \\
&&\hspace{35mm}\Downarrow \Lambda _{GUT}  \notag \\[3mm]
&&\hspace{15mm}SU\left( 3\right) _{C}\otimes SU\left( 2\right) _{L}\otimes
U\left( 1\right) _{Y}  \notag \\[3mm]
&&\hspace{35mm}\Downarrow \Lambda _{EW}  \notag \\[3mm]
&&\hspace{23mm}SU\left( 3\right) _{C}\otimes U\left( 1\right) _{em}  \notag
\end{eqnarray}%
%
%
%
%
%
%
%
%
%
%
%
%
%
%
%
Let us note that among the discrete symmetries we introduced the non-Abelian
flavor symmetry group $T_{7}$, which is the smallest group with a complex
triplet representation, allowing us to naturally accommodate the three
families of fermions.

The fermion assignments under the group 
\mbox{$\mathcal{G}=SU(5)\otimes T_{7}\otimes Z_{2}\otimes Z_{3}\otimes Z_{4}\otimes Z_{4}^{\prime
}\otimes Z_{12}$} are 
\begin{eqnarray}
\Psi _{ij}^{\left( 1\right) } &\sim &\left( \mathbf{10,\mathbf{\mathbf{1}}%
_{0},}1,\omega ,1,1,i\right) ,\hspace{0.5cm}\Psi _{ij}^{\left( 2\right)
}\sim \left( \mathbf{10,\mathbf{1}_{1},}1,\omega ^{2},1,1,e^{\frac{\pi i}{3}%
}\right) ,\hspace{0.5cm}\Psi _{ij}^{\left( 3\right) }\sim \left( \mathbf{10,%
\mathbf{1}_{2},}1,1,1,1,1\right) ,\hspace{3mm}i,j=1,2,3,4,5 \\
\psi ^{i} &=&\left( \psi ^{i\left( 1\right) },\psi ^{i\left( 2\right) },\psi
^{i\left( 3\right) }\right) \sim \left( \overline{\mathbf{5}}\mathbf{,%
\overline{\mathbf{3}},}1,1,1,1,1\right) ,\hspace{0.5cm}\hspace{0.5cm}%
N_{R}\sim \left( \mathbf{1,\mathbf{\mathbf{1}}_{0}},-1,1,1,-1,-1\right) ,%
\hspace{0.5cm}\hspace{0.5cm}\rho \sim \left( \mathbf{24,\mathbf{\mathbf{1}}%
_{0}},-1,1,1,-i,1\right) ,  \notag
\end{eqnarray}%
where $\omega =e^{\frac{2\pi i}{3}}$.

More explicitly, the fermions are accommodated as 
\begin{equation}
\Psi _{ij}^{\left( f\right) }=\frac{1}{\sqrt{2}}\left( 
\begin{array}{ccccc}
0 & u_{3}^{\left( f\right) c} & -u_{2}^{\left( f\right) c} & -u_{1}^{\left(
f\right) } & -d_{1}^{\left( f\right) } \\ 
-u_{3}^{\left( f\right) c} & 0 & u_{1}^{\left( f\right) c} & -u_{2}^{\left(
f\right) } & -d_{2}^{\left( f\right) } \\ 
u_{2}^{\left( f\right) c} & -u_{1}^{\left( f\right) c} & 0 & -u_{3}^{\left(
f\right) } & -d_{3}^{\left( f\right) } \\ 
u_{1}^{\left( f\right) } & u_{2}^{\left( f\right) } & u_{3}^{\left( f\right)
} & 0 & -l^{\left( f\right) c} \\ 
d_{1}^{\left( f\right) } & d_{2}^{\left( f\right) } & d_{3}^{\left( f\right)
} & l^{\left( f\right) c} & 0%
\end{array}%
\right) _{L},\hspace{1.5cm}f=1,2,3\hspace{1.5cm}i,j=1,2,3,4,5.
\end{equation}%
\begin{equation}
\psi ^{i\left( f\right) }=\left( d_{1}^{\left( f\right) c},d_{2}^{\left(
f\right) c},d_{3}^{\left( f\right) c},l^{\left( f\right) },-\nu _{f}\right)
_{L}.
\end{equation}%
Here the subscripts correspond to the different quark colors, while the
superscript $f$ refers to fermion families. One can see that the three
families of left- and right-handed fermions, corresponding to the $\overline{%
\mathbf{5}}$ irrep of $SU(5)$, are unified into a $T_{7}$ antitriplet $%
\overline{\mathbf{3}}$, while the three families of left- and right-handed
fermions corresponding to 
the $\mathbf{10}$ irreps of $SU\left( 5\right) $ are assigned to the three
different $T_{7}$ singlets $\mathbf{\mathbf{\mathbf{1}}_{0},\mathbf{1}_{1},%
\mathbf{1}_{2}}$.

The scalar sector is composed of the following $SU\left( 5\right) $
representations: one $\mathbf{24}$, one $\mathbf{45}$, four $\mathbf{5}$'s
and ten $\mathbf{1}$'s. 
Two sets of $SU\left( 5\right) $ singlets are unified into two $T_{7}$
triplets. The remaining scalar fields, i.e., one $\mathbf{45}$, one $\mathbf{%
24}$, four $\mathbf{5}$'s and the remaining four $\mathbf{1}$'s, are
accommodated by three $T_{7}$ singlets. Thus the $\mathcal{G}$ assignments
of the scalar fields are%
\begin{eqnarray}
\sigma &\sim &\left( \mathbf{1,\mathbf{\mathbf{1}}_{0},}1,1,1,1,e^{-\frac{%
i\pi }{6}}\right) ,\hspace{1cm}\tau \sim \left( \mathbf{1,\mathbf{\mathbf{1}}%
_{0},}1,\omega ,i,1,e^{-\frac{i\pi }{6}}\right) ,\hspace{1cm}\xi =\left( \xi
_{1},\xi _{2},\xi _{3}\right) \sim \left( \mathbf{1,3,-}1,1,1,1,1\right) , 
\notag \\
\varphi &\sim &\left( \mathbf{1,\mathbf{1}_{0},}1,\omega ,-1,1,1\right) ,%
\hspace{0.8cm}\eta \sim \left( \mathbf{1,1,}1,1,1,-1,1\right) ,\hspace{0.8cm}%
\chi =\left( \chi _{1},\chi _{2},\chi _{3}\right) \sim \left( \mathbf{1,3,}%
1,1,1,i,e^{\frac{i\pi }{3}}\right) ,\hspace{0cm}  \notag \\
H_{i}^{\left( 1\right) } &\sim &\left( \mathbf{5,\mathbf{\mathbf{1}}_{0},-}%
1,1,1,1,e^{-\frac{i\pi }{3}}\right) ,\hspace{1cm}H_{i}^{\left( 2\right)
}\sim \left( \mathbf{5,\mathbf{\mathbf{1}}_{0},}1,\omega ,1,1,1\right) ,%
\hspace{1cm}H_{i}^{\left( 3\right) }\sim \left( \mathbf{5,\mathbf{\mathbf{1}}%
_{1},}1,\omega ^{2},1,1,1\right) ,  \notag \\
H_{i}^{\left( 4\right) } &\sim &\left( \mathbf{5,\mathbf{\mathbf{1}}_{2},}%
1,1,1,1,1\right) ,\hspace{1cm}\Xi _{j}^{i}\sim \left( \mathbf{24,\mathbf{%
\mathbf{1}}_{0},}1,1,1,1,1\right) ,\hspace{1cm}\Phi _{jk}^{i}\sim \left( 
\mathbf{45,\mathbf{\mathbf{1}}_{0},-}1,1,1,1,e^{-\frac{i\pi }{3}}\right) .
\end{eqnarray}

The VEVs of the scalars $H_{i}^{\left( h\right) }$ ($h=1,2,3,4$) and $\Xi
_{j}^{i}$ are 
\begin{equation}
\left\langle H_{i}^{\left( h\right) }\right\rangle =v_{H}^{\left( h\right)
}\delta _{i5},\hspace{1cm}h=1,2,3,4,\hspace{1cm}\left\langle \Xi
_{j}^{i}\right\rangle =\frac{2v_{\Xi }}{\sqrt{30}}\,diag\left( 1,1,1,-\frac{3%
}{2},-\frac{3}{2}\right) ,\hspace{1cm}i,j=1,2,3,4,5.
\end{equation}
Note that the VEV pattern for the $\Xi $ field given above is consistent
with the minimization conditions of the model scalar potential and follows
from the general group theory analysis of spontaneous symmetry breakdown 
\cite{Li:1973mq}.

The following comments 
about the possible VEV patterns for the $T_{7}$ scalar triplets $\chi $ and $%
\xi $ are in order. Here we assume a hierarchy between the VEVs of the $%
T_{7} $ scalar triplets $\chi $ and $\xi $, i.e., $v_{\chi }<<v_{\xi }$ ,
which implies that the mixing angle between the the $T_{7}$ scalar triplets $%
\chi $ and $\xi $ is strongly suppressed since it is of the order of $\frac{%
v_{\chi }}{v_{\xi }}$, as follows from the method of recursive expansion of
Refs. \cite{Grimus:2000vj,Alvarado:2012xi,Hernandez:2013mcf}. Consequently,
the mixing between the $T_{7}$ scalar triplets $\chi $ and $\xi $\ can be
neglected. The parts of the scalar potential for each of the two $T_{7}$
scalar triplets at the renormalizable level are given by: 
\begin{eqnarray}
V_{T7}^{\left( 1\right) } &=&-\mu _{\chi }^{2}\left( \chi \chi ^{\ast
}\right) _{\mathbf{\mathbf{\mathbf{1}}_{0}}}+\kappa _{\chi ,1}\left( \chi
\chi ^{\ast }\right) _{\mathbf{3}}\left( \chi \chi ^{\ast }\right) _{%
\overline{\mathbf{3}}}+\kappa _{\chi ,2}\left( \chi \chi \right) _{\mathbf{3}%
}\left( \chi ^{\ast }\chi ^{\ast }\right) _{\overline{\mathbf{3}}}+\kappa
_{\chi ,3}\left( \chi \chi \right) _{\overline{\mathbf{3}}}\left( \chi
^{\ast }\chi ^{\ast }\right) _{\mathbf{3}}  \notag \\
&&+\kappa _{\chi ,4}\left( \chi \chi ^{\ast }\right) _{\mathbf{\mathbf{%
\mathbf{1}}_{0}}}\left( \chi \chi ^{\ast }\right) _{\mathbf{\mathbf{\mathbf{1%
}}_{0}}}+\kappa _{\chi ,5}\left( \chi \chi ^{\ast }\right) _{\mathbf{\mathbf{%
\mathbf{1}}_{1}}}\left( \chi \chi ^{\ast }\right) _{\mathbf{\mathbf{\mathbf{1%
}}_{2}}}+H.c.  \label{T7scalarpotential1}
\end{eqnarray}%
\begin{eqnarray}
V_{T7}^{\left( 2\right) } &=&-\mu _{\xi }^{2}\left( \xi \xi ^{\ast }\right)
_{\mathbf{\mathbf{\mathbf{1}}_{0}}}+\kappa _{\xi ,1}\left( \xi \xi ^{\ast
}\right) _{\mathbf{3}}\left( \xi \xi ^{\ast }\right) _{\overline{\mathbf{3}}%
}+\kappa _{\xi ,2}\left( \xi \xi \right) _{\mathbf{3}}\left( \xi ^{\ast }\xi
^{\ast }\right) _{\overline{\mathbf{3}}}+\kappa _{\xi ,3}\left( \xi ^{\ast
}\xi ^{\ast }\right) _{\mathbf{3}}\left( \xi \xi \right) _{\overline{\mathbf{%
3}}}  \notag \\
&&+\kappa _{\xi ,4}\left( \xi \xi ^{\ast }\right) _{\mathbf{\mathbf{\mathbf{1%
}}_{0}}}\left( \xi \xi ^{\ast }\right) _{\mathbf{\mathbf{\mathbf{1}}_{0}}%
}+\kappa _{\xi ,5}\left( \xi \xi ^{\ast }\right) _{\mathbf{\mathbf{\mathbf{1}%
}_{1}}}\left( \xi \xi ^{\ast }\right) _{\mathbf{\mathbf{\mathbf{1}}_{2}}%
}+H.c.  \label{T7scalarpotential2}
\end{eqnarray}%
In the part of the scalar potential for each $T_{7}$ scalar triplet there
are six free parameters: one bilinear and five quartic couplings. The
minimization conditions of $V_{T7}^{\left( 1\right) }$ and $V_{T7}^{\left(
2\right) }$ lead to the following relations: 
\begin{eqnarray}
\frac{\partial \left\langle V_{T7}^{\left( m\right) }\right\rangle }{%
\partial \text{$v_{S_{1}}$}} &=&-2v_{S_{1}}\mu _{S}+4\kappa _{S,1}\text{$%
v_{S_{1}}$}\left( \text{$v_{S_{2}}^{2}+v_{S_{3}}^{2}$}\right) +4\left(
\kappa _{S,2}+\kappa _{S,3}\right) \left[ 3\text{$v_{S_{1}}^{2}v_{S_{3}}\cos
\left( \theta _{S_{1}}-\theta _{S_{3}}\right) +v_{S_{2}}^{3}$}\cos \left(
\theta _{S_{1}}-\theta _{S_{2}}\right) \right]  \notag \\
&&+8\kappa _{S,4}\text{$v_{S_{1}}\left( \text{$%
v_{S_{1}}^{2}+v_{S_{2}}^{2}+v_{S_{3}}^{2}$}\right) +$}4\kappa _{S,5}\text{$%
v_{S_{1}}\left( 2v_{S_{1}}^{2}\text{$-v_{S_{2}}^{2}-v_{S_{3}}^{2}$}\right) $}
\notag \\
&=&0  \notag \\
\frac{\partial \left\langle V_{T7}^{\left( m\right) }\right\rangle }{%
\partial \text{$v_{S_{2}}$}} &=&-2v_{S_{2}}\mu _{S}+4\kappa _{S,1}\text{$%
v_{S_{2}}$}\left( \text{$v_{S_{1}}^{2}+v_{S_{3}}^{2}$}\right) +4\left(
\kappa _{S,2}+\kappa _{S,3}\right) \left[ 3\text{$v_{S_{2}}^{2}v_{S_{1}}\cos
\left( \theta _{S_{1}}-\theta _{S_{2}}\right) +v_{S_{3}}^{3}$}\cos \left(
\theta _{S_{2}}-\theta _{S_{3}}\right) \right]  \notag \\
&&+8\kappa _{S,4}\text{$v_{S_{2}}\left( \text{$%
v_{S_{1}}^{2}+v_{S_{2}}^{2}+v_{S_{3}}^{2}$}\right) +$}4\kappa _{S,5}\text{$%
v_{S_{2}}\left( 2v_{S_{2}}^{2}\text{$-v_{S_{1}}^{2}-v_{S_{3}}^{2}$}\right) $}
\notag \\
&=&0,  \label{DV} \\
\frac{\partial \left\langle V_{T7}^{\left( m\right) }\right\rangle }{%
\partial \text{$v_{S_{3}}$}} &=&-2v_{S_{3}}\mu _{S}+4\kappa _{S,1}\text{$%
v_{S_{3}}$}\left( \text{$v_{S_{1}}^{2}+v_{S_{2}}^{2}$}\right) +4\left(
\kappa _{S,2}+\kappa _{S,3}\right) \left[ 3\text{$v_{S_{3}}^{2}v_{S_{2}}\cos
\left( \theta _{S_{2}}-\theta _{S_{3}}\right) +v_{S_{1}}^{3}$}\cos \left(
\theta _{S_{1}}-\theta _{S_{3}}\right) \right]  \notag \\
&&+8\kappa _{S,4}\text{$v_{S_{3}}\left( \text{$%
v_{S_{1}}^{2}+v_{S_{2}}^{2}+v_{S_{3}}^{2}$}\right) +$}4\kappa _{S,5}\text{$%
v_{S_{3}}\left( 2v_{S_{3}}^{2}\text{$-v_{S_{1}}^{2}-v_{S_{2}}^{2}$}\right) $}
\notag \\
&=&0,  \notag
\end{eqnarray}%
%
%
%
%
%
%
%
where $m=1,2$, $S=\chi ,\xi $ and $\left\langle S\right\rangle =\left( \text{%
$v_{S_{1}}e^{i\theta _{S_{1}}},v_{S_{2}}e^{i\theta
_{S_{2}}},v_{S_{3}}e^{i\theta _{S_{3}}}$}\right) $. 
Then, from an analysis of the minimization equations given by Eq. (\ref{DV})
and setting $\kappa _{\chi ,2}=-\kappa _{\chi ,3}$, we obtain for a large
range of the parameter space the following VEV direction: 
\begin{equation}
v_{\chi _{1}}=e^{-\frac{i\phi }{2}}\frac{v_{\chi }}{\sqrt{2}},\hspace{0.7cm}%
v_{\chi _{3}}=e^{\frac{i\phi }{2}}\frac{v_{\chi }}{\sqrt{2}},\hspace{0.7cm}%
v_{\chi _{2}}=0,\hspace{0.7cm}v_{\xi _{1}}=v_{\xi _{2}}=v_{\xi _{3}}=\frac{%
v_{\xi }}{\sqrt{3}}.  \label{VEVpattern}
\end{equation}%
%
%
%
%
%
%
%
%
%
%
%
%
%
%
%
In the case of 
$\xi $, this is a vacuum configuration preserving a $Z_{3}$ subgroup of $%
T_{7}$, which has been extensively studied by many authors (see, for
example, Ref. \cite{T7}). The VEV pattern for 
$\chi $ is similar to the one we previously studied in an $SU(5)$ model and
in a 6HDM with $A_{4}$ flavor symmetry \cite{6HDMA4,Campos:2014lla}. It is
worth mentioning that there could be relative phases between the different
components of $\left\langle \xi \right\rangle $, consistent with the scalar
potential minimization equations, as follows from the expressions given in
Eq. (\ref{DV}). We have checked that the nonvanishing phases consistent with
the scalar potential minimization equations satisfy $\theta _{S_{i}}=-\theta
_{S_{j}}\neq \theta _{S_{k}}$, with $i\neq j\neq k$ ($i,j,k=1,2,3$).
Moreover, we checked that the physical observables in the lepton sector,
studied in Sec. \ref{massmix}, do not depend on these phases. 
%
For generality we included nonzero phases in the $\left\langle \chi
\right\rangle $ sector, as indicated in Eq. (\ref{VEVpattern}). 
%
%
%

From the expressions given in Eq. (\ref{DV}), and using the vacuum
configuration for the $T_{7}$ scalar triplets given in Eq. (\ref{VEVpattern}%
), we find the relation between the parameters and the magnitude of the VEV: 
\begin{equation}
\mu _{\chi }^{2}=\left( \kappa _{\chi ,1}+4\kappa _{\chi ,4}+\kappa _{\chi
,5}\right) v_{\chi }^{2},\hspace{0.7cm}\hspace{0.7cm}\mu _{\xi }^{2}=\frac{4%
}{3}\left[ \kappa _{\xi ,1}+2\left( \kappa _{\xi ,2}+\kappa _{\xi ,3}\right)
+3\kappa _{\xi ,4}\right] v_{\xi }^{2}.  \label{mu}
\end{equation}

These results show that the VEV directions for the $T_{7}$ triplets $\chi $
and$\ \xi $ in Eq. (\ref{VEVpattern}) are consistent with a global minimum
of the scalar potential of our model.


%

Assuming that the charged fermion mass pattern and quark mixing hierarchy is
caused by the $Z_{3}$, $Z_{4}$, and $Z_{12}$ symmetries, and in order to
relate the quark masses with the quark mixing parameters, we set the VEVs of
the $SU(5)$ scalar singlets as follows: 
\begin{equation}
v_{\chi }<<v_{\eta }\sim v_{\varphi }=v_{\tau }=v_{\xi }=v_{\sigma }=\Lambda
_{GUT}=\lambda \Lambda ,  \label{sizeVEVsinglets}
\end{equation}%
where $\lambda =0.225$ is one of the parameters in the Wolfenstein
parametrization and $\Lambda $ is the high-energy scale cutoff of our model,
to be clarified below. 
Assuming that the parameters of the scalar interaction terms involving these 
$SU(5)$ scalar singlets are of the same order of magnitude, it is
straightforward to show that the VEVs in Eq. (\ref{sizeVEVsinglets}) 
are consistent with the minimization conditions of the model scalar
potential. 

The fields $\Phi _{jk}^{i}$, being the $\mathbf{45}$ irrep of $SU(5)$,
satisfy the following relations: 
\begin{equation}
\Phi _{jk}^{i}=-\Phi _{kj}^{i},\hspace{1.5cm}\sum_{i=1}^{5}\Phi _{ij}^{i}=0,%
\hspace{1.5cm}i,j,k=1,2,\cdots ,5.
\end{equation}%
%
%
%
%
%
%
%
%
%
%
%
%
%
%
%
Consequently, the only allowed nonzero VEVs of $\Phi _{jk}^{i}$ are 
\begin{equation}
\left\langle \Phi _{p5}^{p}\right\rangle =-\frac{1}{3}\left\langle \Phi
_{45}^{4}\right\rangle =v_{\Phi },\hspace{1.5cm}\left\langle \Phi
_{j5}^{i}\right\rangle =v_{\Phi }\left( \delta _{j}^{i}-4\delta
_{4}^{i}\delta _{j}^{4}\right) ,\hspace{1.5cm}i,j=1,2,3,4,5,\hspace{1.5cm}%
p=1,2,3,5.
\end{equation}%
%
%
%
%
%
%
%
%
%
%
%
%
%
%
%
With the specified particle content, there are the following interaction
terms, invariant under the group $\mathcal{G}$ and relevant for the further
analysis: 
\begin{eqnarray}
\tciLaplace _{Y} &=&\alpha _{1}\left( \psi ^{i}\xi \right) _{\mathbf{\mathbf{%
1}}_{0}}H^{j\left( 1\right) }\Psi _{ij}^{\left( 1\right) }\frac{\sigma
^{5}\varphi ^{2}+\kappa \sigma \tau ^{4}\varphi ^{\ast 2}}{\Lambda ^{8}}%
+\alpha _{2}\left( \psi ^{i}\xi \right) _{\mathbf{1}_{2}}H^{j\left( 1\right)
}\Psi _{ij}^{\left( 2\right) }\frac{\tau ^{4}}{\Lambda ^{5}}+\alpha
_{3}\left( \psi ^{i}\xi \right) _{\mathbf{1}_{1}}H^{j\left( 1\right) }\Psi
_{ij}^{\left( 3\right) }\frac{\sigma ^{2}}{\Lambda ^{3}}  \notag \\
&&+\beta _{1}\left( \psi ^{i}\xi \right) _{\mathbf{\mathbf{1}}_{0}}\Phi
_{i}^{jk}\Psi _{jk}^{\left( 1\right) }\frac{\sigma ^{5}\varphi ^{2}+\kappa
\sigma \tau ^{4}\varphi ^{\ast 2}}{\Lambda ^{8}}+\beta _{2}\left( \psi
^{i}\xi \right) _{\mathbf{1}_{2}}\Phi _{i}^{jk}\Psi _{jk}^{\left( 2\right) }%
\frac{\tau ^{4}}{\Lambda ^{5}}+\beta _{3}\left( \psi ^{i}\xi \right) _{%
\mathbf{1}_{1}}\Phi _{i}^{jk}\Psi _{jk}^{\left( 3\right) }\frac{\sigma ^{2}}{%
\Lambda ^{3}}  \notag \\
&&+\varepsilon ^{ijklp}\left\{ \gamma _{11}\Psi _{ij}^{\left( 1\right)
}H_{p}^{\left( 2\right) }\Psi _{kl}^{\left( 1\right) }\frac{\sigma ^{6}}{%
\Lambda ^{6}}+\gamma _{12}\Psi _{ij}^{\left( 1\right) }H_{p}^{\left(
4\right) }\Psi _{kl}^{\left( 2\right) }\frac{\sigma ^{5}}{\Lambda ^{5}}%
+\gamma _{21}\Psi _{ij}^{\left( 2\right) }H_{p}^{\left( 4\right) }\Psi
_{kl}^{\left( 1\right) }\frac{\sigma ^{5}}{\Lambda ^{5}}\right.  \notag \\
&&+\left. \gamma _{22}\Psi _{ij}^{\left( 2\right) }H_{p}^{\left( 3\right)
}\Psi _{kl}^{\left( 2\right) }\frac{\sigma ^{4}}{\Lambda ^{4}}+\gamma
_{13}\Psi _{ij}^{\left( 1\right) }H_{p}^{\left( 3\right) }\Psi _{kl}^{\left(
3\right) }\frac{\sigma ^{3}}{\Lambda ^{3}}+\gamma _{31}\Psi _{ij}^{\left(
3\right) }H_{p}^{\left( 3\right) }\Psi _{kl}^{\left( 1\right) }\frac{\sigma
^{3}}{\Lambda ^{3}}\right.  \notag \\
&&+\left. \gamma _{23}\Psi _{ij}^{\left( 2\right) }H_{p}^{\left( 2\right)
}\Psi _{kl}^{\left( 3\right) }\frac{\sigma ^{2}}{\Lambda ^{2}}+\gamma
_{32}\Psi _{ij}^{\left( 3\right) }H_{p}^{\left( 2\right) }\Psi _{kl}^{\left(
2\right) }\frac{\sigma ^{2}}{\Lambda ^{2}}+\gamma _{33}\Psi _{ij}^{\left(
3\right) }H_{p}^{\left( 4\right) }\Psi _{kl}^{\left( 3\right) }\right\} 
\notag \\
&&+\frac{\lambda _{1\nu }}{\Lambda ^{2}}\left[ \psi ^{i}\left( \chi ^{\ast
}\chi ^{\ast }\right) _{\mathbf{3}}\right] _{\mathbf{\mathbf{1}}%
_{0}}H_{i}^{\left( 1\right) }N_{R}+\frac{\lambda _{2\nu }}{\Lambda ^{2}}%
\left[ \left( \psi ^{i}\chi ^{\ast }\right) _{\mathbf{3}}\chi ^{\ast }\right]
_{\mathbf{\mathbf{1}}_{0}}H_{i}^{\left( 1\right) }N_{R}+\frac{\lambda _{3\nu
}}{\Lambda }\left( \psi ^{i}\chi \right) _{\mathbf{\mathbf{1}}%
_{0}}H_{j}^{\left( 1\right) }\rho _{i}^{j}  \notag \\
&&+\frac{\lambda _{4\nu }}{\Lambda }\left( \psi ^{i}\chi \right) _{\mathbf{%
\mathbf{1}}_{0}}\Phi _{ij}^{k}\rho _{k}^{j}+m_{N}\overline{N}%
_{R}N_{R}^{c}+y_{1}\overline{N}_{R}N_{R}^{c}\frac{\sigma ^{\ast }\sigma
+x_{1}\tau ^{\ast }\tau +x_{2}\varphi ^{\ast }\varphi }{\Lambda }%
+y_{2}Tr\left( \rho ^{2}\right) \eta +y_{3}Tr\left( \rho ^{2}\Xi \right) 
\frac{\eta }{\Lambda },  \label{LY}
\end{eqnarray}

where the dimensionless couplings in Eq. (\ref{LY}) are $\mathcal{O}(1)$
parameters, and are assumed to be real, excepting $\gamma _{11}$, $\gamma
_{12}$, $\gamma _{21}$, $\gamma _{31}$ and $\gamma _{13}$, which are assumed
to be complex. The subscripts $\mathbf{\mathbf{1}}_{0}\mathbf{,\mathbf{1}%
_{1},\mathbf{1}_{2}}$ denote projecting out the corresponding $T_{7}$
singlet in the product of the two triplets. Let us note that Eq. (\ref{LY})
is $SU\left( 5\right) $ invariant, since the scalar fields $H_{p}^{\left(
h\right) }$, $H^{\left( h\right) p}$ ($h=1,2,3,4$), $\Phi _{jk}^{i}$, $\Phi
_{i}^{jk}$\ transform as $\mathbf{5}$, $\overline{\mathbf{5}}$, $\mathbf{45}$
and $\overline{\mathbf{45}}$ under $SU\left( 5\right) $, respectively, and
the fermionic fields $\psi ^{i}$ and $\Psi _{ij}^{\left( f\right) }$ ($%
f=1,2,3$) as $\overline{\mathbf{5}}$ and $\mathbf{10}$, under the $SU\left(
5\right) $ group, respectively. Besides that, it is worth mentioning that
the scalar field $H^{\left( h\right) p}$ ($h=1,2,3,4$)\ transforms with the
opposite $Z_{N}$ charges as compared to $H_{p}^{\left( h\right) }$.
Furthermore, it is noteworthy that the term $Tr\left( \rho ^{2}\right) $ is
not present in Eq. (\ref{LY}), since this term is not invariant under the $%
Z_{4}^{\prime }$ symmetry. The lightest of the physical neutral scalar
states of $H^{(1)}$, $H^{(2)}$, $H^{(3)}$, $H^{(4)}$ and $\Phi $, should be
interpreted as the SM-like 126 GeV Higgs observed at the LHC.%

Let us summarize and comment on the above presented model setup. 
In comparison with the next-to-minimal $SU(5)$ GUT model of Refs. \cite%
{Georgi:1979df,Frampton:1979,Ellis:1979,Nandi:1980sd,Frampton:1980,Langacker:1980js,Kalyniak:1982pt,Giveon:1991,Dorsner:2007fy,Dorsner:2006dj,FileviezPerez:2007nh,Perez:2008ry,Khalil:2013ixa}%
, besides for the introduction of additional discrete symmetries, we also
extended the fermionic sector by introducing one heavy Majorana neutrino $%
N_{R}$, singlet under the SM group and a $\mathbf{24}$ fermionic irrep of $%
SU\left( 5\right) $, namely $\rho _{j}^{i}$. 
%
We will show that since the $Z_{2}$ symmetry present in (\ref{Symmetry-1})
is not preserved at low energies, the active neutrinos get tree-level masses
via type-I and type-III seesaw mechanisms. In the next section, we will also
show that in order to successfully accommodate the experimental data on
neutrino mass-squared splittings, one needs both the SM singlet right-handed
Majorana neutrino and the $\mathbf{24}$ fermionic irrep of $SU(5)$. Having
only one of them would lead to two massless active neutrinos, in
contradiction to the experimental data on neutrino oscillations. 
Note that our fermionic sector is less minimal than the one considered in
Ref. \cite{Perez:2007rm} with only a $\mathbf{24}$ fermionic irrep of $%
SU\left( 5\right) $. However, our model provides a successful description of
the SM charged fermion masses and mixing pattern, not addressed in Ref.\cite%
{Perez:2007rm}.

Despite the flavor-discrete groups in Eq. (\ref{Symmetry-1}), the
corresponding field assignment as well as the VEV pattern look rather
sophisticated, although each introduced element plays its own role in the
arrangement of the desired particle spectrum and flavor mixing. Let as
briefly sketch our justification of the model setup: %
%
%

\begin{enumerate}
\item The scalar sector includes the following $SU\left( 5\right) $
representations: one $\mathbf{24}$, one $\mathbf{45}$, four $\mathbf{5}$'s
and ten $\mathbf{1}$'s. The $\mathbf{45}$ and the four $\mathbf{5}$'s scalar
irreps of $SU\left( 5\right) $ acquire VEVs at the electroweak scale, thus
inducing the second step of symmetry breaking. The remaining scalars acquire
VEVs at the GUT scale and trigger the first step of symmetry breaking. As
previously mentioned, having scalar fields in the $\mathbf{45}$
representation of $SU\left( 5\right) $ is crucial to get the correct mass
relations of down-type quarks and charged leptons.


\item The $T_{7}$ discrete group is crucial to generating textures for the
lepton sector that successfully account for the experimentally observed
deviation from the trimaximal mixing pattern that attracted a lot of
attention in the literature as a framework for describing the lepton
mixings; see for example Ref.~\cite{T7}. 
To reproduce the nontrivial quark mixing consistent with experimental data,
the up-type quark sector requires three $\mathbf{5}$'s, i.e., $H_{i}^{\left(
2\right) }$, $H_{i}^{\left( 3\right) }$, and $H_{i}^{\left( 4\right) }$
irreps of $SU(5)$ assigned to different $T_{7}$ singlets. In the down-type
quark sector, on the other hand, only one $\mathbf{5}$ irrep $H_{i}^{\left(
1\right) }$, one $\mathbf{45}$ irrep $\Phi _{jk}^{i}$ assigned to $T_{7}$
trivial singlets\ and three $\mathbf{\mathbf{1}}$'s, unified in the $T_{7}$
triplet $\xi $, are needed.

\item The $Z_{2}$ symmetry separates the scalars in the $\mathbf{5}$ and $%
\mathbf{45}$ irreps of $SU\left( 5\right) $ participating in the Yukawa
interactions for charged leptons and down-type quarks from those ones
participating in the Yukawa interactions for up-type quarks. This implies
that the $SU\left( 5\right) $ scalar multiplets contributing to the masses
of the down-type quarks and charged leptons are different from those that
provide masses to the up-type quarks. Furthermore, the $Z_{2}$ symmetry
separates the $T_{7}$ scalar triplet $\xi $ participating in the Yukawa
interactions for charged leptons and down-type quarks, from that one ($\chi $%
) participating in the neutrino Yukawa interactions. In the scalar sector,
the $Z_{2}$ symmetry distinguishes the $T_{7}$\ scalar triplet $\xi $, the $%
SU(5)$ multiplets $H_{i}^{\left( 1\right) }$ and $\Phi _{jk}^{i}$ charged
under this symmetry, from the remaining scalar fields, neutral under this
symmetry. Because of this, the $\mathbf{45}$ and one of the $\mathbf{5}$'s
scalars participate in the Yukawa interactions for leptons and down-type
quarks, whereas the remaining $SU\left( 5\right) $ multiplets participate in
the Yukawa interactions for up-type quarks. 
This results in a reduction of parameters in the quark sector, since due to
the $Z_{2}$ symmetry the $\mathbf{45}$ scalar irrep of $SU\left( 5\right) $
does not appear in the up-type quark Yukawa terms. 
Furthermore, all fermions are $Z_{2}$ even excepting the right-handed
Majorana neutrino and the $\mathbf{24}$ fermionic irrep of $SU\left(
5\right) $, which are $Z_{2}$ odd.

\item As with the $T_{7}$ symmetry, the $Z_{4}^{\prime }$ symmetry is also
necessary to get a predictive neutrino mass-matrix texture that only depends
on three effective parameters and that gives rise to the experimentally
observed deviation from the trimaximal mixing pattern. This symmetry also
separates the $\mathbf{24}$ fermionic irrep $\rho $ and the charged under
this symmetry from the remaining fermionic fields, neutral under this
symmetry.

\item The $Z_{3}$ and $Z_{4}$ symmetries are crucial to getting the right
pattern of charged lepton and down-type quark masses. The $Z_{3}$ symmetry
distinguishes the three $\mathbf{10}$'s irreps of $SU(5)$, having different $%
Z_{3}$ charges. The $Z_{4}$ symmetry separates the $SU(5)$ scalar singlets $%
\varphi $ and $\tau $, charged under this symmetry, from the remaining
scalar fields, neutral under this symmetry. All the fermionic fields are
assumed to transform trivially under the $Z_{4}$ symmetry. 
Without the $Z_{3}$ and $Z_{4}$ symmetries, the down quark and electron
masses would be larger by about 2 orders of magnitude than their
corresponding experimental values, unless one sets the corresponding Yukawa
couplings unnaturally small. It is noteworthy that, unlike in the up-type
quark sector, a $\lambda ^{8}$ suppression ($\lambda =0.225$ is one of the
Wolfenstein parameters) in the 11 entry of the mass matrices for down-type
quarks and charged leptons is required to naturally explain the smallness of
the down quark and electron masses. The $Z_{3}$, $Z_{4}$, and $Z_{12}$
symmetries will be crucial to achieve that $\lambda ^{8}$ suppression. %

\item The $Z_{12}$ symmetry shapes the hierarchical structure of the quark
mass matrices necessary to get a realistic pattern of quark masses and
mixings. Besides that, the charged lepton mass hierarchy also arises from
the $Z_{12}$ symmetry. Let us recall that due to the properties of the $%
Z_{N} $ groups, it follows that $Z_{12}$ is the lowest cyclic symmetry that
allows building the dimension-ten up-type quark Yukawa term with a $\sigma
^{6}/\Lambda ^{6}$ insertion in a term of dimension four, crucial to getting
the required $\lambda ^{6}$ suppression in the 11 entry of the up-type quark
mass matrix. 
\end{enumerate}


\section{Lepton masses and mixing}

\label{massmix} The charged lepton mass matrix is derived from Eq. (\ref{LY}%
) by using the product rules for the $T_{7}$ group given in Appendix \ref{A}%
, and considering that the VEV pattern of the $SU(5)$ singlet $T_{7}$ scalar
triplet $\xi $ satisfies Eq. (\ref{VEVpattern}) with the VEVs of their
components set to be equal to $\lambda \Lambda $ ($\Lambda $ being the
cutoff of our model), as indicated by Eq. (\ref{sizeVEVsinglets}). 
Then, the mass matrix for charged leptons takes the form 
\begin{equation}
M_{l}=\frac{v}{\sqrt{2}}V_{lL}^{\dag }\left( 
\begin{array}{ccc}
a_{1}^{\left( l\right) }\lambda ^{8} & 0 & 0 \\ 
0 & a_{2}^{\left( l\right) }\lambda ^{5} & 0 \\ 
0 & 0 & a_{3}^{\left( l\right) }\lambda ^{3}%
\end{array}%
\right) =V_{lL}^{\dag }\mathrm{diag}\left( m_{e},m_{\mu },m_{\tau }\right) ,
\label{Lepton-Mass-Matrix-1-1}
\end{equation}%
with 
\begin{eqnarray}
a_{1}^{\left( l\right) } &=&\frac{1}{v}\left( \alpha _{1}v_{H}^{(1)}-6\beta
_{1}v_{\Phi }\right) ,\hspace{1cm}a_{2}^{\left( l\right) }=\frac{1}{v}\left(
\alpha _{2}v_{H}^{(1)}-6\beta _{2}v_{\Phi }\right) ,\hspace{1cm}%
a_{3}^{\left( l\right) }=\frac{1}{v}\left( \alpha _{3}v_{H}^{(1)}-6\beta
_{3}v_{\Phi }\right) ,  \label{yl} \\
V_{lL} &=&\frac{1}{\sqrt{3}}\left( 
\begin{array}{ccc}
1 & 1 & 1 \\ 
1 & \omega & \omega ^{2} \\ 
1 & \omega ^{2} & \omega%
\end{array}%
\right) ,\hspace{2cm}\omega =e^{\frac{2\pi i}{3}}.  \label{yl2}
\end{eqnarray}%
Here $\lambda =0.225$ is the Wolfenstein parameter. As was commented in the
previous section, we assume that the dimensionless couplings $\alpha _{i}$
and $\beta _{i}$ ($i=1,2,3$) in Eq. (\ref{LY}) are roughly of the same order
of magnitude and the VEVs $v_{H}^{(1)}$ and $v_{\Phi }$ are of the order of
the electroweak scale $v\simeq 246$ GeV. Therefore, the hierarchy among the
charged lepton masses arises from the breaking of the $Z_{3}$, $Z_{4}$ and $%
Z_{12}$ symmetries. As seen, the lepton mass matrix Eq. (\ref%
{Lepton-Mass-Matrix-1-1}) is fully determined in our model by three
effective parameters $\alpha _{1,2,3}^{(l)}$ shown in Eq. (\ref{yl}), which
we fit to reproduce the experimentally measured values of lepton masses and
mixings at the $M_{Z}$ scale. A similar situation takes place in the sector
of quarks, as will be shown in the next section.

From the neutrino Yukawa terms of Eq. (\ref{LY}), and taking into account
that the VEVs of the $SU(5)$ singlets $\varphi $, $\sigma $ and $\tau $ are
set to be equal to $\lambda \Lambda $ (where $\Lambda $ is our model
cutoff), as indicated by Eq. (\ref{sizeVEVsinglets}), 
we find that the fields contained in the $\mathbf{24}$ fermionic irrep of $%
SU(5)$ acquire very large masses, which are given by 
\begin{eqnarray}
m_{\rho _{0}} &=&\allowbreak y_{2}v_{\eta }-\frac{y_{3}v_{\Xi }v_{\eta }}{%
\sqrt{30}\Lambda },  \notag  \label{M24-1} \\
m_{\rho _{3}} &=&\allowbreak \allowbreak y_{2}v_{\eta }-\frac{3y_{3}v_{\Xi
}v_{\eta }}{\sqrt{30}\Lambda },  \notag \\
m_{\rho _{8}} &=&\allowbreak \allowbreak y_{2}v_{\eta }+\frac{2y_{3}v_{\Xi
}v_{\eta }}{\sqrt{30}\Lambda },  \notag \\
m_{\rho _{_{\left( 3,2\right) }}} &=&m_{\rho _{\left( \overline{3},2\right)
}}=\allowbreak \allowbreak y_{2}v_{\eta }-\frac{y_{3}v_{\Xi }v_{\eta }}{2%
\sqrt{30}\Lambda }.
\end{eqnarray}%
Here $m_{\rho _{0}}$, $m_{\rho _{3}}$ and $m_{\rho _{8}}$ are the masses of
the fermionic singlet $\rho _{0}$, triplet $\rho _{3}$ and octet $\rho _{8}$
contained in the $\mathbf{24}$ fermionic irrep of $SU\left( 5\right) $,
respectively. 
We denote by $m_{\rho _{\left( 3,2\right) }}$ and $m_{\rho _{\left( 
\overline{3},2\right) }}$ the masses of the $\left( 3,2\right) $ and $\left( 
\overline{3},2\right) $ fermionic fields corresponding to the $SU(3)$
triplet and $SU(3)$ antitriplet, $SU(2)$ doublet parts of $\rho $,
respectively. Consequently, the light active neutrino masses arise from
type-I and type-III seesaw mechanisms induced by the $SU(5)$ singlet heavy
Majorana neutrino $N_{R}$, the fermionic singlet $\rho _{0}$ and the
fermionic triplet $\rho _{3}$, respectively.

From the neutrino Yukawa terms of Eq. (\ref{LY}) and the VEV pattern of the $%
SU(5)$ singlet $T_{7}$ scalar triplet $\chi $ given by Eq. (\ref{VEVpattern}%
), 
we find the neutrino mass matrix: 
\begin{eqnarray}
M_{\nu } &=&\left( 
\begin{array}{cc}
O_{3\times 3} & M_{\nu }^{D} \\ 
\left( M_{\nu }^{D}\right) ^{T} & M_{R}%
\end{array}%
\right) ,\hspace{2cm}M_{\nu }^{D}=\allowbreak \left( 
\begin{array}{ccccc}
0 & Y_{1}e^{-\frac{i\phi }{2}} & Y_{2}e^{-\frac{i\phi }{2}} & Y_{2}e^{-\frac{%
i\phi }{2}} & Y_{2}e^{-\frac{i\phi }{2}} \\ 
X & 0 & 0 & 0 & 0 \\ 
0 & Y_{1}e^{\frac{i\phi }{2}} & Y_{2}e^{\frac{i\phi }{2}} & Y_{2}e^{\frac{%
i\phi }{2}} & Y_{2}e^{\frac{i\phi }{2}}%
\end{array}%
\right) ,\hspace{2cm}  \notag \\
M_{R} &=&\left( 
\begin{array}{ccccc}
m_{N} & 0 & 0 & 0 & 0 \\ 
0 & m_{\rho _{0}} & 0 & 0 & 0 \\ 
0 & 0 & m_{\rho _{3}} & 0 & 0 \\ 
0 & 0 & 0 & m_{\rho _{3}} & 0 \\ 
0 & 0 & 0 & 0 & m_{\rho _{3}}%
\end{array}%
\right) ,
\end{eqnarray}%
where: 
\begin{equation}
X=\left( \lambda _{1\nu }+\lambda _{2\nu }\right) v_{H}^{\left( 1\right) }%
\frac{v_{\chi }^{2}}{\Lambda ^{2}},\hspace{1cm}Y_{1}=\frac{\sqrt{15}}{2}%
\left( \frac{1}{5}\lambda _{3\nu }v_{H}^{(1)}+\lambda _{4\nu }v_{\Phi
}\right) \frac{v_{\chi }}{\Lambda },\hspace{1cm}Y_{2}=\left( \lambda _{3\nu
}v_{H}^{(1)}-3\lambda _{4\nu }v_{\Phi }\right) \frac{v_{\chi }}{\Lambda },
\label{Y-def-1}
\end{equation}%
Therefore, the light neutrino mass matrix takes the following form: 
\begin{equation}
M_{L}=M_{\nu }^{D}M_{R}^{-1}\left( M_{\nu }^{D}\right) ^{T}\allowbreak
=\left( 
\begin{array}{ccc}
Ae^{-i\phi } & 0 & A \\ 
0 & B & 0 \\ 
A & 0 & Ae^{i\phi }%
\end{array}%
\right) \allowbreak ,  \label{ML-1}
\end{equation}%
where 
\begin{equation}
A=\allowbreak \frac{Y_{1}^{2}}{m_{\rho _{0}}}+\frac{3Y_{2}^{2}}{m_{\rho _{3}}%
},\hspace{2cm}B=\frac{X^{2}}{m_{N}}.  \label{Aeq}
\end{equation}%
%
%
%
%
%
%
%
%
%
%
%
%
The smallness of neutrino masses in our model is the consequence of their
inverse scaling with respect to the large masses of the singlet $\rho _{0}$
and the triplet $\rho _{3}$ fermionic fields and proportionality to the
squared neutrino Yukawa couplings.

The mass matrix $M_{L}$ in Eq. (\ref{ML-1}) for light active neutrinos is
diagonalized by a unitary rotation matrix $V_{\nu }$. There are two
solutions of this diagonalization problem: 
\begin{eqnarray}
V_{\nu }^{\dagger }M_{L}(V_{\nu }^{\dagger })^{T} &=&\left( 
\begin{array}{ccc}
m_{1} & 0 & 0 \\ 
0 & m_{2} & 0 \\ 
0 & 0 & m_{3}%
\end{array}%
\right) ,\hspace{0.5cm}\mbox{with}\hspace{0.5cm}V_{\nu }=\left( 
\begin{array}{ccc}
\cos \theta & 0 & \sin \theta e^{-i\phi } \\ 
0 & 1 & 0 \\ 
-\sin \theta e^{i\phi } & 0 & \cos \theta%
\end{array}%
\right) P_{\nu },\hspace{0.5cm}\theta =\pm \frac{\pi }{4},  \label{Vnu-11} \\
P_{\nu } &=&diag\left( e^{i\alpha _{1}/2},e^{i\alpha _{2}/2},e^{i\alpha
_{3}/2}\right) .  \label{Vnu}
\end{eqnarray}%
The solutions corresponding to $\theta =+\pi /4$ and $\theta =-\pi /4$ we
identify 
with the normal (NH) and inverted (IH) neutrino mass hierarchies,
respectively, so that 
\begin{eqnarray}
\mbox{NH} &:&\theta =+\frac{\pi }{4}:\hspace{10mm}m_{\nu _{1}}=0,\hspace{10mm%
}m_{\nu _{2}}=B,\hspace{10mm}m_{\nu _{3}}=2A,\hspace{10mm}\alpha _{1}=\alpha
_{2}=0,\hspace{10mm}\alpha _{3}=\phi ,  \label{mass-spectrum-Inverted} \\%
[0.12in]
\mbox{IH} &:&\theta =-\frac{\pi }{4}:\hspace{10mm}m_{\nu _{1}}=2A,\hspace{8mm%
}m_{\nu _{2}}=B,\hspace{10mm}m_{\nu _{3}}=0,\hspace{12.5mm}\alpha
_{2}=\alpha _{3}=0,\hspace{10mm}\alpha _{1}=-\phi .
\label{mass-spectrum-Normal}
\end{eqnarray}%
Let us note the presence of nonvanishing Majorana phases $\phi $ and $-\phi $
for NH and IH cases, respectively. This simple relation requires the
effective dimensionful parameters $A$ and $B$, given by Eq. (\ref{Aeq}), to
be real, which is consistent with our previously mentioned assumption that
the dimensionless couplings in Eq. (\ref{LY}) are $\mathcal{O}(1)$
parameters assumed to be real, excepting $\gamma _{11}$, $\gamma _{12}$, $%
\gamma _{21}$, $\gamma _{31}$ and $\gamma _{13}$, assumed to be complex 

Now we find the Pontecorvo-Maki-Nakagawa-Sakata (PMNS) leptonic mixing
matrix, 
\begin{equation}
U=V_{lL}^{\dag }V_{\nu }=\left( 
\begin{array}{ccc}
\frac{\cos \theta }{\sqrt{3}}-\frac{e^{i\phi }\sin \theta }{\sqrt{3}} & 
\frac{1}{\sqrt{3}} & \frac{\cos \theta }{\sqrt{3}}+\frac{e^{-i\phi }\sin
\theta }{\sqrt{3}} \\ 
&  &  \\ 
\frac{\cos \theta }{\sqrt{3}}-\frac{e^{i\phi +\frac{2i\pi }{3}}\sin \theta }{%
\sqrt{3}} & \frac{e^{-\frac{2i\pi }{3}}}{\sqrt{3}} & \frac{e^{\frac{2i\pi }{3%
}}\cos \theta }{\sqrt{3}}+\frac{e^{-i\phi }\sin \theta }{\sqrt{3}} \\ 
&  &  \\ 
\frac{\cos \theta }{\sqrt{3}}-\frac{e^{i\phi -\frac{2i\pi }{3}}\sin \theta }{%
\sqrt{3}} & \frac{e^{\frac{2i\pi }{3}}}{\sqrt{3}} & \frac{e^{-\frac{2i\pi }{3%
}}\cos \theta }{\sqrt{3}}+\frac{e^{-i\phi }\sin \theta }{\sqrt{3}}%
\end{array}%
\right) P_{\nu },  \label{PMNS}
\end{equation}%
%
%
%
%
%
%
with the pattern of the trimaximal (TM$_{2}$) type \cite{TM}. 
It is noteworthy that in our model the PMNS matrix depends on a single
parameter $\phi $, and the neutrino masses (\ref{mass-spectrum-Inverted})
and (\ref{mass-spectrum-Normal}) depend on two parameters, $A$ and $B$. 
Comparing the matrix $U$ in Eq. (\ref{PMNS}) with the standard
parameterization of the PMNS matrix in terms of the solar $\theta _{12}$,
the atmospheric $\theta _{23}$ and the reactor $\theta _{13}$ angles, we
find 
\begin{eqnarray}
&&\sin ^{2}\theta _{12}=\frac{\left\vert U_{e2}\right\vert ^{2}}{%
1-\left\vert U_{e3}\right\vert ^{2}}=\frac{1}{2-z\cdot \cos \phi },\hspace{%
20mm}  \notag \\[3mm]
&&\sin ^{2}\theta _{13}=\left\vert U_{e3}\right\vert ^{2}=\frac{1}{3}%
(1+z\cdot \cos \phi ),  \notag \\
&&\sin ^{2}\theta _{23}=\frac{\left\vert U_{\mu 3}\right\vert ^{2}}{%
1-\left\vert U_{e3}\right\vert ^{2}}=\frac{2-z\cdot (\cos \phi +\sqrt{3}\sin
\phi )}{4-2z\cdot \cos \phi },  \label{theta-ij}
\end{eqnarray}%
with $z=1$ and $z=-1$ for NH and IH, respectively. Note that 
in the limit $\phi =0$ and $\phi =\pi $ for IH and NH, respectively, the
mixing matrix in Eq. (\ref{PMNS}) reduces to 
the tribimaximal mixing pattern, which yields a vanishing reactor mixing
angle $\theta _{13}$. 

For the Jarlskog invariant and the CP-violating phase \cite{PDG}, we find 
\begin{equation}
J=\func{Im}\left( U_{e1}U_{\mu 2}U_{e2}^{\ast }U_{\mu 1}^{\ast }\right) =-%
\frac{1}{6\sqrt{3}}\cos 2\theta =0,\hspace{2cm}\sin \delta =\frac{8J}{\cos
\theta _{13}\sin 2\theta _{12}\sin 2\theta _{23}\sin 2\theta _{13}}=0
\end{equation}%
since 
$\theta =\pm \frac{\pi }{4}$ according to Eq. (\ref{Vnu-11}). 

Thus, our model predicts a vanishing leptonic Dirac CP-violating phase. 

\quad 
In what follows, we adjust the three free effective parameters $\phi $, $A$
and $B$ of the active neutrino sector of our model to reproduce the
experimental values of 
three leptonic mixing parameters and two neutrino mass-squared splittings,
reported in Tables \ref{NH} and \ref{IH}, for the normal and inverted
neutrino mass hierarchies, respectively. We fit the parameter $\phi $ to
adjust the experimental values of the leptonic mixing parameters $\sin
^{2}\theta _{ij}$, whereas $A$ and $B$ are fixed so that the measured
mass-squared differences are reproduced for the normal (NH) and inverted
(IH) neutrino mass hierarchies. From Eqs. (\ref{mass-spectrum-Normal}), (\ref%
{mass-spectrum-Inverted}), and the definition $\Delta
m_{ij}^{2}=m_{i}^{2}-m_{j}^{2}$, we find 
\begin{eqnarray}
&&\mbox{NH}:\ m_{\nu _{1}}=0,\ \ \ m_{\nu _{2}}=B=\sqrt{\Delta m_{21}^{2}}%
\approx 9\mbox{meV},\ \ \ m_{\nu _{3}}=2\left\vert A\right\vert =\sqrt{%
\Delta m_{31}^{2}}\approx 50\mbox{meV};  \label{AB-Delta-IH} \\[0.12in]
&&\mbox{IH}\hspace{2mm}:\ m_{\nu _{2}}=B=\sqrt{\Delta m_{21}^{2}+\Delta
m_{13}^{2}}\approx 50\mbox{meV},\ \ \ \ \ m_{\nu _{1}}=2\left\vert
A\right\vert =\sqrt{\Delta m_{13}^{2}}\approx 49\mbox{meV},\ \ \ m_{\nu
_{3}}=0,  \label{AB-Delta-NH}
\end{eqnarray}%
for the best-fit values of $\Delta m_{ij}^{2}$ taken from Tables \ref{NH}
and \ref{IH}. 

\quad 
To fit the leptonic mixing parameters $\sin ^{2}\theta _{ij}$ to their
experimental values, given in Tables \ref{NH}, \ref{IH}, we vary the $\phi $
parameter, finding the following best fit result: 
\begin{eqnarray}
&&\mbox{NH}\ :\ \phi =-0.88\pi ,\ \ \ \sin ^{2}\theta _{12}\approx 0.34,\ \
\ \sin ^{2}\theta _{23}\approx 0.61,\ \ \ \sin ^{2}\theta _{13}\approx
0.0232;  \label{parameter-fit-IN} \\[0.12in]
&&\mbox{IH}\hspace{2.5mm}:\ \phi =\ \ 0.12\,\pi ,\ \ \ \ \ \sin ^{2}\theta
_{12}\approx 0.34,\ \ \ \sin ^{2}\theta _{23}\approx 0.61,\ \ \ \ \,\sin
^{2}\theta _{13}\approx 0.0238.  \label{parameter-fit-IH}
\end{eqnarray}

Thus, $\sin ^{2}\theta _{13}$ is in excellent agreement with the
experimental data, for both normal and inverted neutrino mass hierarchies,
whereas $\sin ^{2}\theta _{12}$ and $\sin ^{2}\theta _{23}$ exhibit a $%
2\sigma $ 
deviation from their best-fit values. 

Now we are ready to make a prediction for the neutrinoless double beta ($%
0\nu \beta \beta $) decay amplitude, which is proportional to the effective
Majorana neutrino mass parameter, 
\begin{equation}
m_{\beta \beta }=\biggl|\sum_{k}U_{ek}^{2}m_{\nu _{k}}\biggr|,  \label{mee}
\end{equation}%
where $U_{ek}^{2}$ and $m_{\nu _{k}}$ are the PMNS mixing matrix elements
and the Majorana neutrino masses, respectively. Using Eqs. (\ref{Vnu-11})-(%
\ref{PMNS}) and (\ref{AB-Delta-IH})-(\ref{parameter-fit-IH}), we get 
for both normal and inverted hierarchies 
\begin{equation}
m_{\beta \beta }=\frac{1}{3}\left( B+4A\cos ^{2}\frac{\phi }{2}\right)
=\left\{ 
\begin{array}{l}
4\ \mbox{meV}\ \ \ \ \ \ \ \mbox{for \ \ \ \ NH} \\ 
50\ \mbox{meV}\ \ \ \ \ \ \ \mbox{for \ \ \ \ IH} \\ 
\end{array}%
\right.  \label{eff-mass-pred}
\end{equation}%
%
%
%
%
%
%
%
%
%
%
%
%
%
%
%
%
These values are beyond the reach of the present and forthcoming $0\nu \beta
\beta $ decay experiments. The presently best upper limit on the effective
neutrino mass is $m_{\beta \beta }\leq 160$ meV, which arises from the
recently quoted EXO-200 experiment \cite{Auger:2012ar} $T_{1/2}^{0\nu \beta
\beta }(^{136}\mathrm{Xe})\geq 1.6\times 10^{25}$ yr at 90\% C.L. This limit
will be improved within a not too distant future. The GERDA
\textquotedblleft phase-II\textquotedblright experiment \cite%
{Abt:2004yk,Ackermann:2012xja} 
is expected to reach 
\mbox{$T^{0\nu\beta\beta}_{1/2}(^{76}{\rm Ge}) \geq
2\times 10^{26}$ yr}, corresponding to $m_{\beta \beta }\leq 100$ meV. A
bolometric CUORE experiment, using ${}^{130}Te$ \cite{Alessandria:2011rc},
is currently under construction. It has an estimated sensitivity around $%
T_{1/2}^{0\nu \beta \beta }(^{130}\mathrm{Te})\sim 10^{26}$ yr, which
corresponds to \mbox{$m_{\beta\beta}\leq 50$ meV.} There are also proposals
for ton-scale next-to-next generation $0\nu \beta \beta $ experiments with $%
^{136}$Xe \cite{KamLANDZen:2012aa,Auty:2013:zz} and $^{76}$Ge \cite%
{Abt:2004yk,Guiseppe:2011me} claiming sensitivities over $T_{1/2}^{0\nu
\beta \beta }\sim 10^{27}$ yr, corresponding to $m_{\beta \beta }\sim 12-30$
meV. For recent reviews, see for example Ref. \cite{Reviewbetadecay}.
Consequently, as it can be seen from Eq. (\ref{eff-mass-pred}), our model
predicts $T_{1/2}^{0\nu \beta \beta }$ at the level of sensitivities of the
next generation or next-to-next generation $0\nu \beta \beta $ experiments.

\section{Quark masses and mixing}

\label{Quarkmixing} Using Eq. (\ref{LY}), together with the product rules
for the $T_{7}$ group given in Appendix \ref{A}, and considering that the
components of the $SU(5)$ singlet $T_{7}$ scalar triplet $\xi $ acquire the
same VEV as shown by Eq. (\ref{VEVpattern}), which is set to be equal to $%
\lambda \Lambda $ as the VEV of the $Z_{12}$ charged scalar $\sigma $ as
indicated by Eq. (\ref{sizeVEVsinglets}) ($\Lambda $ being the cutoff of our
model), we find the quark mass matrices:%
\begin{equation}
M_{U}=\left( 
\begin{array}{ccc}
a_{11}^{\left( U\right) }\lambda ^{6} & a_{12}^{\left( U\right) }\lambda ^{5}
& a_{13}^{\left( U\right) }\lambda ^{3} \\ 
a_{12}^{\left( U\right) }\lambda ^{5} & a_{22}^{\left( U\right) }\lambda ^{4}
& a_{23}^{\left( U\right) }\lambda ^{2} \\ 
a_{13}^{\left( U\right) }\lambda ^{3} & a_{23}^{\left( U\right) }\lambda ^{2}
& a_{33}^{\left( U\right) }%
\end{array}%
\right) \allowbreak \frac{v}{\sqrt{2}},  \label{MU}
\end{equation}%
\begin{equation}
M_{D}=\frac{v}{\sqrt{2}}\left( 
\begin{array}{ccc}
a_{1}^{\left( D\right) }\lambda ^{8} & 0 & 0 \\ 
0 & a_{2}^{\left( D\right) }\lambda ^{5} & 0 \\ 
0 & 0 & a_{3}^{\left( D\right) }\lambda ^{3}%
\end{array}%
\right) \left( V_{lL}^{\dag }\right) ^{T}=diag\left(
m_{d},m_{s},m_{b}\right) \left( V_{lL}^{\dag }\right) ^{T},  \label{MD}
\end{equation}%
where $\lambda =0.225$ and the $\mathcal{O}(1)$ dimensionless couplings in
Eqs. (\ref{MU}) and (\ref{MD}) are given by:

\begin{eqnarray}
a_{12}^{\left( U\right) } &=&2\sqrt{2}\left( \gamma _{12}+\gamma
_{21}\right) \frac{v_{H}^{\left( 4\right) }}{v},\hspace{1cm}a_{11}^{\left(
U\right) }=4\sqrt{2}\gamma _{11}\frac{v_{H}^{\left( 2\right) }}{v},\hspace{%
1cm}a_{13}^{\left( U\right) }=2\sqrt{2}\left( \gamma _{13}+\gamma
_{31}\right) \frac{v_{H}^{\left( 3\right) }}{v},  \notag \\
a_{23}^{\left( U\right) } &=&2\sqrt{2}\left( \gamma _{23}+\gamma
_{32}\right) \frac{v_{H}^{\left( 2\right) }}{v},\hspace{1cm}a_{22}^{\left(
U\right) }=4\sqrt{2}\gamma _{22}\frac{v_{H}^{\left( 3\right) }}{v},\hspace{%
1cm}a_{33}^{\left( U\right) }=4\sqrt{2}\gamma _{33}\frac{v_{H}^{\left(
4\right) }}{v},  \notag \\
a_{1}^{\left( D\right) } &=&\frac{1}{v}\left( \alpha _{1}v_{H}^{(1)}+2\beta
_{1}v_{\Phi }\right) ,\hspace{1cm}a_{2}^{\left( D\right) }=\frac{1}{v}\left(
\alpha _{2}v_{H}^{(1)}+2\beta _{2}v_{\Phi }\right) ,\hspace{1cm}%
a_{3}^{\left( D\right) }=\frac{1}{v}\left( \alpha _{3}v_{H}^{(1)}+2\beta
_{3}v_{\Phi }\right) .
\end{eqnarray}%
From Eq. (\ref{MD}) it follows that the CKM quark mixing mixing matrix does
not receive contributions from the down-type quark sector, meaning that
quark mixing arises solely from the up-type quark sector. Consequently,
having a realistic up-type quark masses and quark mixing angles requires
that the mass matrix for up-type quarks to be given by: 
\begin{equation}
M_{U}=V_{CKM}diag\left( m_{u},m_{c},m_{t}\right) V_{CKM}^{T}.
\end{equation}

Using the standard parametrization for the CKM quark mixing matrix, together
with the relations $m_{u}=a\lambda ^{8}m_{t}$ and $m_{c}=b\lambda ^{4}m_{t}$%
, where $a$ and $b\mathcal{\ }$ are $\mathcal{O}(1)$ coefficients, we get
the mass matrix for up-type quarks described in Eq. (\ref{MU}), with entries
exhibiting different scalings in terms of powers of the Wolfenstein
parameter $\lambda =0.225$. Thus, from the requirement of realistic up quark
masses and quark mixing angles with $\mathcal{O}(1)$ dimensionless couplings
in Eq. (\ref{MU}), we find a $\lambda ^{6}$ suppression in the 11 entry of
the up-quark mass matrix instead of a $\lambda ^{8}$ one. We have
numerically checked that a $\lambda ^{6}$ supression in the 11 entry of the
up-quark mass matrix, with $\mathcal{O}(1)$ dimensionless couplings, is
consistent with realistic up-quark masses and quark mixing angles.

Assuming that the quark 
mass and mixing pattern is caused by the breaking of the $Z_{3}$, $Z_{4}$
and $Z_{12}$ symmetries, to simplify our analysis, we adopt 
a benchmark where the dimensionless charged fermion Yukawa couplings are
approximately equal. Specifically, we set 
\begin{eqnarray}
\gamma _{11} &=&\left( 1-\frac{\lambda ^{2}}{2}\right) ^{1/2}\gamma
_{1}e^{i\phi _{1}},\hspace{1cm}\gamma _{12}=\gamma _{21}=-\gamma
_{1}e^{i\phi _{2}},\hspace{1cm}\gamma _{22}=\gamma _{1}\left( 1-\frac{%
\lambda ^{2}}{2}\right) ^{-1/2},\hspace{1cm}  \notag \\
\gamma _{13} &=&\gamma _{31}=\gamma _{2}\left( 1-\frac{\lambda ^{2}}{2}%
\right) ^{3}e^{i\phi _{3}},\hspace{1cm}\gamma _{23}=\gamma _{32}=-\gamma
_{2},\hspace{1cm}\alpha _{i}=\beta _{i},\hspace{1cm}i=1,2,3,
\label{apuniversality}
\end{eqnarray}%
with $\gamma _{1}$, $\gamma _{2}$, $\alpha _{i}$ and $\beta _{i}$ ($i=1,2,3$%
)\ real $\mathcal{O}(1)$ parameters. Our benchmark of nearly equal charged
fermion Yukawa couplings given in Eq. (\ref{apuniversality}), which we have
numerically checked is consistent with realistic up-quark masses and quark
mixing angles, is also adopted in order to reduce the number of free
effective parameters in the quark sector of our model. There is no tuning in
the parameters of our model. Let us note that the exactly equal
dimensionless quark Yukawa couplings do not allow generating the up and
charm quark masses. In Appendix \ref{sec:YC-univ} we give another possible
motivation for the approximate universality of dimensionless couplings,
which can be studied beyond the present model, adding new symmetries. 
\begin{table}[tbh]
\begin{center}
\begin{tabular}{c|l|l}
\hline\hline
Observable & Model value & Experimental value \\ \hline
$m_{u}(MeV)$ & \quad $0.86$ & \quad $1.45_{-0.45}^{+0.56}$ \\ \hline
$m_{c}(MeV)$ & \quad $673$ & \quad $635\pm 86$ \\ \hline
$m_{t}(GeV)$ & \quad $174.2$ & \quad $172.1\pm 0.6\pm 0.9$ \\ \hline
$m_{d}(MeV)$ & \quad $2.9$ & \quad $2.9_{-0.4}^{+0.5}$ \\ \hline
$m_{s}(MeV)$ & \quad $57.7$ & \quad $57.7_{-15.7}^{+16.8}$ \\ \hline
$m_{b}(GeV)$ & \quad $2.82$ & \quad $2.82_{-0.04}^{+0.09}$ \\ \hline
$\bigl|V_{ud}\bigr|$ & \quad $0.974$ & \quad $0.97427\pm 0.00015$ \\ \hline
$\bigl|V_{us}\bigr|$ & \quad $0.2257$ & \quad $0.22534\pm 0.00065$ \\ \hline
$\bigl|V_{ub}\bigr|$ & \quad $0.00305$ & \quad $%
0.00351_{-0.00014}^{+0.00015} $ \\ \hline
$\bigl|V_{cd}\bigr|$ & \quad $0.2256$ & \quad $0.22520\pm 0.00065$ \\ \hline
$\bigl|V_{cs}\bigr|$ & \quad $0.97347$ & \quad $0.97344\pm 0.00016$ \\ \hline
$\bigl|V_{cb}\bigr|$ & \quad $0.0384$ & \quad $0.0412_{-0.0005}^{+0.0011}$
\\ \hline
$\bigl|V_{td}\bigr|$ & \quad $0.00785$ & \quad $%
0.00867_{-0.00031}^{+0.00029} $ \\ \hline
$\bigl|V_{ts}\bigr|$ & \quad $0.0377$ & \quad $0.0404_{-0.0005}^{+0.0011}$
\\ \hline
$\bigl|V_{tb}\bigr|$ & \quad $0.999145$ & \quad $%
0.999146_{-0.000046}^{+0.000021}$ \\ \hline
$J$ & \quad $2.32\times 10^{-5}$ & \quad $(2.96_{-0.16}^{+0.20})\times
10^{-5}$ \\ \hline
$\delta $ & \quad $64^{\circ }$ & \quad $68^{\circ }$ \\ \hline\hline
\end{tabular}%
\end{center}
\caption{Model and experimental values of the quark masses and CKM
parameters.}
\label{Observables}
\end{table}
%

Besides that, for simplicity we assume that the complex phase responsible
for CP violation in the quark sector arises solely from the up-type quark
sector, as indicated by Eq. (\ref{apuniversality}). In addition, to simplify
the analysis, we fix $a_{33}^{\left( U\right) }=1$, as suggested by
naturalness arguments. Consequently, the up-type quark mass matrix reads 
\begin{equation}
M_{U}=\left( 
\begin{array}{ccc}
a_{1}^{\left( U\right) }\left( 1-\frac{\lambda ^{2}}{2}\right) ^{1/2}\lambda
^{6}e^{i\phi _{1}} & -a_{1}^{\left( U\right) }\lambda ^{5}e^{i\phi _{2}} & 
a_{2}^{\left( U\right) }\left( 1-\frac{\lambda ^{2}}{2}\right) ^{3}\lambda
^{3}e^{i\phi _{3}} \\ 
-a_{1}^{\left( U\right) }\lambda ^{5}e^{i\phi _{2}} & a_{1}^{\left( U\right)
}\left( 1-\frac{\lambda ^{2}}{2}\right) ^{-1/2}\lambda ^{4} & -a_{2}^{\left(
U\right) }\lambda ^{2} \\ 
a_{2}^{\left( U\right) }\left( 1-\frac{\lambda ^{2}}{2}\right) ^{3}\lambda
^{3}e^{i\phi _{3}} & -a_{2}^{\left( U\right) }\lambda ^{2} & 1%
\end{array}%
\right) \allowbreak \frac{v}{\sqrt{2}},
\end{equation}%
As seen from the above formulas, the quark sector of our model contains ten
parameters, i.e, $\lambda $, $a_{33}^{\left( U\right) }$, $a_{1}^{\left(
U\right) }$, $a_{2}^{\left( U\right) }$, $a_{1}^{\left( D\right) }$, $%
a_{2}^{\left( D\right) }$, $a_{3}^{\left( D\right) }$ and the phases $\phi
_{l}$ ($l=1,2,3)$, to describe the quark mass and mixing pattern, which is
characterized by ten physical observables, i.e., the six quark masses, the
three mixing angles and the CP phase. 
Out of the ten model parameters two of them, $\lambda $ and $a_{33}^{\left(
U\right) }$, are fixed, whereas the remaining eight are fitted to reproduce
the six quark masses and four quark mixing parameters. In 
Table \ref{Observables} we show the experimental values of the physical
observables in the quark sector, together with our results obtained for the
following best-fit values of the model parameters: 
\begin{eqnarray}
a_{1}^{\left( U\right) } &\simeq &1.96,\hspace{1cm}a_{2}^{\left( U\right)
}\simeq 0.74,\hspace{1cm}\phi _{1}\simeq 10.94^{\circ },\hspace{1cm}\phi
_{2}\simeq 6.02^{\circ },\hspace{1cm}\phi _{3}\simeq 21.65^{\circ },\hspace{%
1cm}  \notag \\
a_{1}^{\left( D\right) } &\simeq &2.54,\hspace{1cm}a_{2}^{\left( D\right)
}\simeq 0.58,\hspace{1cm}a_{3}^{\left( D\right) }\simeq 1.42.
\end{eqnarray}%
%
%
We use the experimental values for the quark masses at the $M_{Z}$ scale,
reported in Ref. (\cite{Bora:2012tx}) (which are similar to those in \cite%
{Xing:2007fb}), whereas the experimental values of the CKM matrix elements,
the Jarlskog invariant $J$ and the CP-violating phase $\delta $ are taken
from Ref. \cite{PDG}. 
%
Let us note that the agreement of our model with the experimental data is as
good as in the models of Refs. \cite%
{Branco2010,CarcamoHernandez:2010im,Hernandez:2013hea,King:2013hj,Hernandez:2014vta,Campos2014,Hernandez:2015dga,CarcamoHernandez2015}%
, and better than in Refs. \cite%
{Chen:2007,Xing2010,Branco2012,CarcamoHernandez:2012xy,Bhattacharyya:2012pi,Vien:2014ica,KhalilDelta27,Ishimori:2014jwa,Ishimori:2014nxa}%
. %
The following comparison of our model with these models could be in order. 
Despite the similar quality of the data description our model is more
predictive than the model of Ref. \cite{Branco2010}, since the latter, %
focused only on the quark sector, has a total of 12 free parameters, whereas
the quark sector of our model is described by 8 free effective parameters
that are adjusted to reproduce the 10 physical observables of the quark
sector. %
The models of Ref.\cite{King:2013hj}, Ref.\cite%
{CarcamoHernandez:2010im,KhalilDelta27}, Ref. \cite{Xing2010}, Refs.\cite%
{Ishimori:2014jwa,Ishimori:2014nxa,CarcamoHernandez2015}, Refs. \cite%
{CarcamoHernandez:2012xy,Campos2014,Hernandez:2015dga,Hernandez:2013hea,Hernandez:2014vta}%
, Refs. \cite{Bhattacharyya:2012pi,Vien:2014ica} and Ref. \cite{Branco2012} 
possess in the quark sector 
6, 7, 8, 9, 10, 12, and 13 free parameters. 
The total number of the effective free parameters of our $T_{7}$ flavor
adjoint $SU(5)$ GUT model 
is 16, 
from which 2 are fixed and 14 are fitted to reproduce the experimental
values of 18 observables in the quark and lepton sectors. On the other hand
the $SU(5)$ model with $T^{\prime }\otimes Z_{12}\otimes Z_{12}^{\prime }$
symmetry of Ref. \cite{Chen:2007} has nine parameters in the Yukawa sector
for the charged fermions and the neutrinos. 
However it does not 
account for CP violation in the quark sector, whereas our model does.
Furthermore, 
the values of the physical observables we derived in our model for both
quark and lepton sectors exhibit a significantly better agreement with their
corresponding experimental values than those derived in Ref. \cite{Chen:2007}
within the $SU(5)$ model with $T^{\prime }\otimes Z_{12}\otimes
Z_{12}^{\prime }$ symmetry. 

\section{Gauge coupling $SU(5)$ unification}

\label{Unification}

In the previous sections, we analyzed the possibility of describing the quark
and lepton masses and flavor mixing within the framework of our model. This
analysis was based on the symmetries of the model and particular field
assignments to the symmetry group representations, which allowed us to
single out several effective parameters completely determining the lepton
and quark mass matrices. It is a notable property of the model that the SM
fermion mass matrices depend on the ``fundamental'' parameters of the model
Lagrangian only through this set of a few effective parameters. Now we turn
to more subtle aspects of the model depending on the details of the non-SM
components of the $SU(5)$ multiplets, as well as on the ``fundamental''
parameters, which may have crucial impact on its ultraviolet behavior. 

As is well known, there are many extensions of the SM which lead to gauge
coupling unification (GCU) and also successfully fulfill all the constraints
coming from fermion masses, proton decay, and perturbativity. In particular,
models based on supersymmetric and also nonsupersymmetric (non-SUSY) $SU(5)$
unification have been widely studied in the literature
 \cite%
{FileviezPerez:2007nh,Senjanovic:2009kr}. For non-SUSY $SU(5)$ scenarios,
the unification of gauge couplings can be as good as or better than in the
MSSM, 
despite the fact that the number of extra fields beyond the SM is 
smaller. More restrictive 
conditions such as 
the possibility of implementation of an appropriate neutrino mass generation
mechanism, and compatibility with 
the existing phenomenological, cosmological and astrophysical constraints,
require some specific properties for this extra field content. As was
already pointed out in Sec. II, the $SU(5)$ scalar representations with
the minimal number of Higgs bosons needed to generate the fermions masses
and mixings are one $\mathbf{24}_{s}$, one $\mathbf{45}_{s}$, four $\mathbf{%
5}_{s}$'s, and twelve $\mathbf{1}_{s}$'s. If an extra fermionic $\mathbf{24}%
_{f}$ representation is also 
included, a 
simple configuration of the extra fields allowing 
the type-I and type-III seesaw mechanisms of neutrino mass generation 
can be constructed with masses below the GUT scale. 
As will be shown later the condition of the GCU and compatibility with 
the lower experimental limit on the proton decay half-life predicts the
masses for the extra scalar fields within the LHC reach. This opens up the
possibility for experimental tests of the considered models. 
%
On the other hand, 
the dark matter constraints on the scalar sector will lead us to the
conclusion that the masses for the type-I and type-III fermionic seesaw
mediators, $m_{_{NR}}$, $m_{\rho_{0}}$ and $m_{\rho_{3}}$, 
should be 
at high energies far beyond the TeV ballpark.

It is noteworthy to mention that our configurations rely on fine-tuning
which separate light from heavy degrees of freedom. However, this issue is
an unavoidable problem of grand unified theories. Evidence of this is the
standard $SU(5)$, which suffers from what is known as the doublet-triplet
splitting problem i.e, the fact that the SM Higgs doublet has a mass of $%
m_{h}\sim 125$ GeV while the colored triplet in the 5 irrep must have a mass
of the order of the GUT scale in order to prevent proton decay. Although
several solutions to this problem have been suggested in the literature (for
a short review, see Ref. \cite{Randall:1995sh}), we do not consider here any
particular one. Instead, we view this as a fine-tuning problem and, in
fact, to make the exotic particles in our model light will require, in
general, additional fine-tunings. We assured ourselves that the large number
of free and uncorrelated parameters in the scalar potential will allow us to
reproduce the required mass differences. There are non-SUSY $SU(5)$ models
which also heavily rely on fine-tuning among states, and which do not
consider formal solutions to this problem, as seen in Refs. \cite%
{Emmanuel-Costa:2013gia,FileviezPerez:2007nh}. We are aware of this fine-
tuning problem, whose formal solution goes beyond the scope of this work and
is deferred for a future publication.

\subsubsection{Setup}

\label{sec:Setup} The scalar sector of our model is composed, as described
in the previous sections, by the $\mathbf{5}_{s}$, $\mathbf{24}_{s}$, and $%
\mathbf{45}_{s}$ irreps of $SU(5)$ with the following $SU(3)_{C}\times
SU(2)_{L}\times U(1)_{Y}$ decompositions: 
\begin{eqnarray}  \label{GUT1}
\mathbf{5}_{s} &=& (1,2)_{1/2}\oplus (3,1)_{-1/3},  \notag \\
&=& H_{1}\oplus H_{2},  \notag \\
\mathbf{24}_{s} &=& (1,1)_{0}\oplus (1,3)_{0} \oplus (8,1)_{0}\oplus
(3,2)_{-5/6} \oplus (\overline{3},2)_{5/6},  \notag \\
&=& \Xi_{1} \oplus \Xi_{3} \oplus \Xi_{8} \oplus \Xi_{(3,2)} \oplus \Xi_{(%
\overline{3},2)},  \notag \\
\mathbf{45}_{s} &=& (1,2)_{1/2}\oplus (3,1)_{-1/3} \oplus (3,3)_{-1/3}
\oplus (\overline{3},1)_{4/3}\oplus (\overline{3},2)_{-7/6} \oplus
(6,1)_{-1/3} \oplus (8,2)_{1/2},  \notag \\
&=& \Phi_{1} \oplus \Phi_{2} \oplus \Phi_{3} \oplus \Phi_{4} \oplus \Phi_{5}
\oplus \Phi_{6} \oplus \Phi_{7}.  \notag \\
\end{eqnarray}
The SM Higgs doublet is embedded in the $\mathbf{5}_{s}$. The adjoint $%
\mathbf{24}_{s}$ representation triggers the breaking of $SU(5)$ to the SM
at the GUT scale. As we previously commented, this particle content should be
further extended in order to account for the nonzero neutrino masses. 
We introduced a fermionic $\mathbf{24}_{f}$ irrep of the $SU(5)$. 
It has the SM decomposition 
\begin{align}
\mathbf{24}_{f} & = (1,1)_{0} \oplus (1,3)_{0} \oplus (8,1)_{0} \oplus
(3,2)_{-5/6} \oplus (\overline{3},2)_{5/6},  \notag \\
& = \rho_{0} \oplus \rho_{3} \oplus \rho_{8} \oplus \rho_{(3,2)} \oplus
\rho_{(\overline{3},2)}.  \label{GUT2}
\end{align}
%
In this case the $SU(5)$ singlet $N_{R}$ Majorana neutrino with the heavy SM
singlet $\rho_{0}$ mediate type-I seesaw mechanism while the $SU(2)_{L}$ triplet $\rho_{3}$ 
%
gives rise to the type-III seesaw mechanism. 
%
Combining these three extra fermions, $N_{R}$, $\rho_{0}$, and $\rho_{3}$,
with some of the scalar fields from the $\mathbf{5}_{s}$, $\mathbf{24}_{s}$,
and $\mathbf{45}_{s}$ irreps, we construct the simplest benchmark
configurations that unify the gauge couplings within the $SU(5)$ and satisfy
some general requirements in order to guarantee their phenomenological
viability. These requirements are\newline
\textbf{(i)} \textit{Perturbative $SU(5)$ unification}: This means that
gauge couplings unify as well as or even better than in the MSSM, and the
value of $\alpha_{G}$ is in the perturbative regime. Note that we are not
necessarily imposing the exact unification of the gauge couplings at the GUT
scale ($m_{G}$). Rather, we 
allow for a difference of the gauge couplings at $m_{G}$ falling into the
area of the MSSM "nonunification triangle"
 \cite%
{DeRomeri:2011ie,Arbelaez:2013hr}.\newline
\textbf{(ii)} \textit{Proton decay}: There are some specific fields
contributing to proton decay. The dimension-six proton decay operators are
mediated by the superheavy gauge bosons, usually named leptoquarks, in the $%
\mathbf{24}$ irrep: $\rho_{(3,2)} \oplus \rho_{(\overline{3},2)}=(3,2,-5/6)
\oplus (\overline{3} ,2,5/6)$, which must be heavier than $3\times 10^{15}$ GeV to satisfy the experimental lower bound on the proton decay lifetime. 
Here, we assume that these fields live at the GUT scale. In addition, we
deal with field configurations which, in almost all the parameter space,
fulfill the current constraint from $\tau_{p\rightarrow \pi^{0}e^{+}}
\gtrsim 10^{34}$  years \cite{Nath:2006ut,Abe:2013lua}. This, through the
relation for the proton decay half-life, $\Gamma =
\alpha_{G}^{2}m_{p}^{5}/m_{G}^{4}$, leads to a GUT scale 
of $m_{G}\gtrsim 3\times 10^{15}$  GeV.\newline
\textbf{(iii)} \textit{Fermion masses}: In particular, neutrino mass
generation through the type-I and type-III seesaw mechanisms  \cite%
{Perez:2007rm,Biggio:2010me}. The configurations should then contain at
least one copy of the fermionic fields $N_{R}$, $\rho_{0}$, and $\rho_{3}$,
as described before.\newline
\textbf{(iv)} \textit{Nontrivial phenomenology}: Among the models passed
through the above three conditions, we select those models which may provide
a distinguishing signal at the LHC.

In what follows, we analyze some of the "minimal" benchmark models, which
lead to correct unification and also fulfill the above listed requirements,
with as simple a field content as possible. As the preferable models, we
consider those which are phenomenologically rich enough, in the sense that
some of the fields (being colored) could give rise to certain resonances at
the LHC. The analyzed models will lead to an available parameter space for
the different masses of the scalars and the type-I, type-III seesaw fermionic
mediators.


\subsubsection{The Models}

\label{sec:The Models} For simplicity, and following the notations of Ref. 
\cite{Perez:2008ry}, we rewrite the masses of the 
fermionic $\mathbf{24}_{f}$ 
given in Eq. (\ref{M24-1}) as follows: 
\begin{align}
m_{\rho _{3}}& =m-3e^{i\alpha }\Lambda ^{^{\prime }},  \notag \\
m_{\rho _{8}}& =m+2e^{i\alpha }\Lambda ^{^{\prime }},  \notag \\
m_{\rho _{_{\left( 3,2\right) }}}& =m_{\rho _{\left( \overline{3},2\right)
}}=m-1/2e^{i\alpha }\Lambda ^{^{\prime }},  \notag \\
m_{\rho _{0}}& =m-e^{i\alpha }\Lambda ^{^{\prime }}.  \label{GUT3}
\end{align}%
where $\alpha $ is the relative phase between $y_{2}$ and $y_{3}$. For the
particular case where $\alpha =0$, the parameters $\Lambda ^{^{\prime }}$
and $m$ are then defined as $m=y_{2}v_{\eta }$ and $\Lambda ^{^{\prime }}=%
\hat{y}_{3}v_{\Xi }\frac{v_{\eta }}{\Lambda }$ (where $\hat{y}_{3}=y_{3}/%
\sqrt{30}$). 

In order to deal with the simplest models, we look for configurations where
the only contribution from the fermionic $\mathbf{24}_{f}$ to the RGE flow 
comes from the fermionic type-I and type-III seesaw mediators. We assume the
other $\mathbf{24}_{f}$ components, $\rho_{8}$, $\rho_{(3,1)}$, and $\rho_{(%
\overline{3},2)}$, have no RGE effect, being as heavy as the GUT scale. It
is worth reiterating that in the analyzed benchmark models the GCU is
achieved having a few particles with masses below the GUT scale: the
fermions $N_{R}$, $\rho_{0}$, $\rho_{3}$ plus some of the extra scalar
fields from the $\mathbf{5}_{s}$, $\mathbf{24}_{s}$, and $\mathbf{45}_{s}$
multiplets. This kind of spectra can be easily obtained by fine-tuning Eq. (%
\ref{GUT3}), imposing that the mass of the remaining fermionic fields in the 
$\mathbf{24}_{f}$ lives at the GUT scale or above.

Searching for the models, we keep our analysis at the one-loop level. It
could be easily extended to two loops. 
However, this sophistication makes no 
impact on 
our final results and conclusions.

\begin{figure}[htb]
\caption{The parameter space passing the conditions \textbf{(i)}-\textbf{(iv)%
} in Sec. \protect\ref{sec:Setup}. The region compatible with the proton
decay constraints $m_{G}\geq 3\times 10^{15}$GeV is explicitly shown in
light blue. The upper two panels correspond to \textit{Model (1)} and the
lower one to \textit{Model (2)} considered in the text. In the latter case,
the mass of the extra scalar electroweak doublet $\Phi_{1}$ is fixed at $2$
TeV. }
\label{GUT6}\centering
\begin{tabular}{cc}
\includegraphics[width=0.42\textwidth]{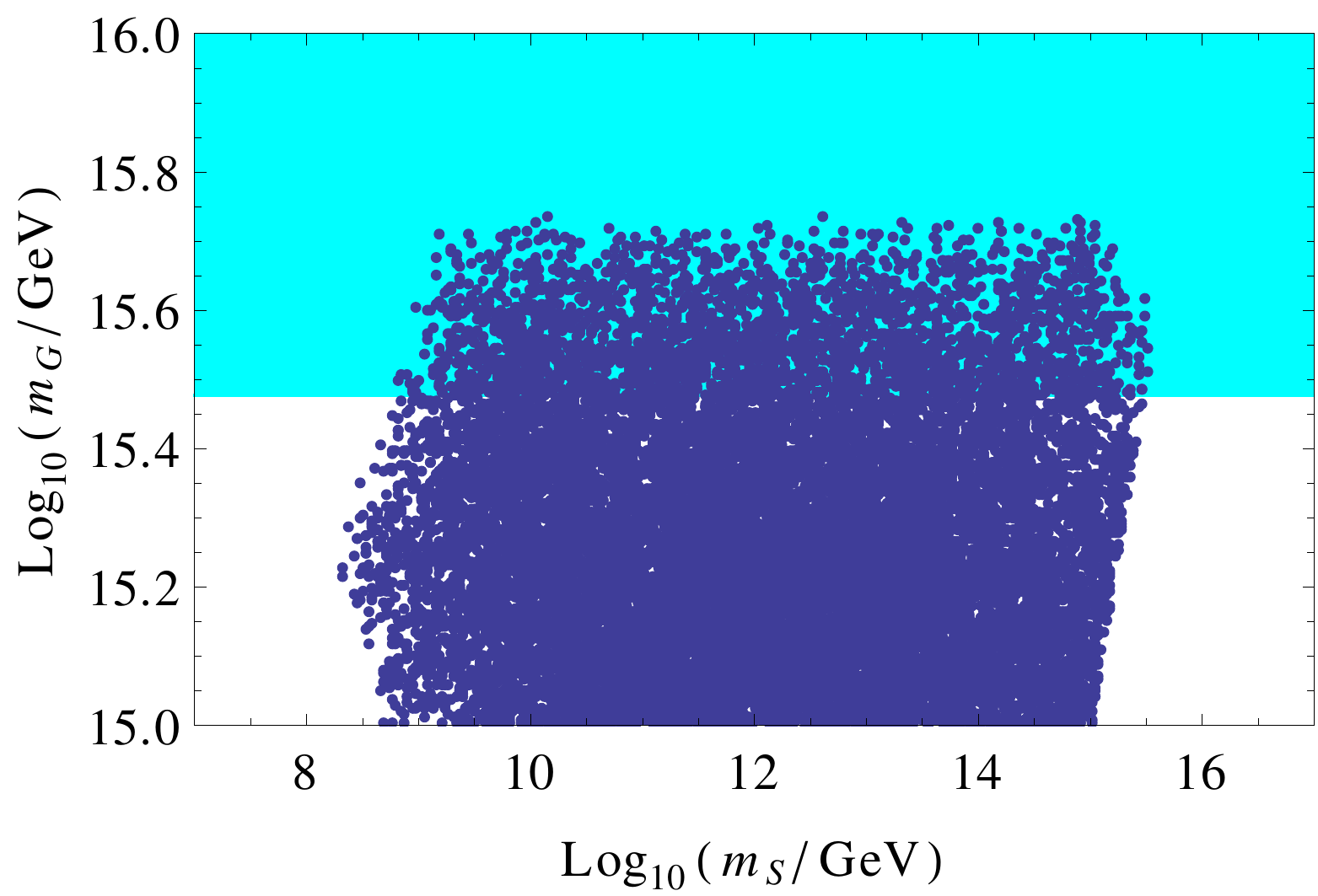} & %
\includegraphics[width=0.42\textwidth]{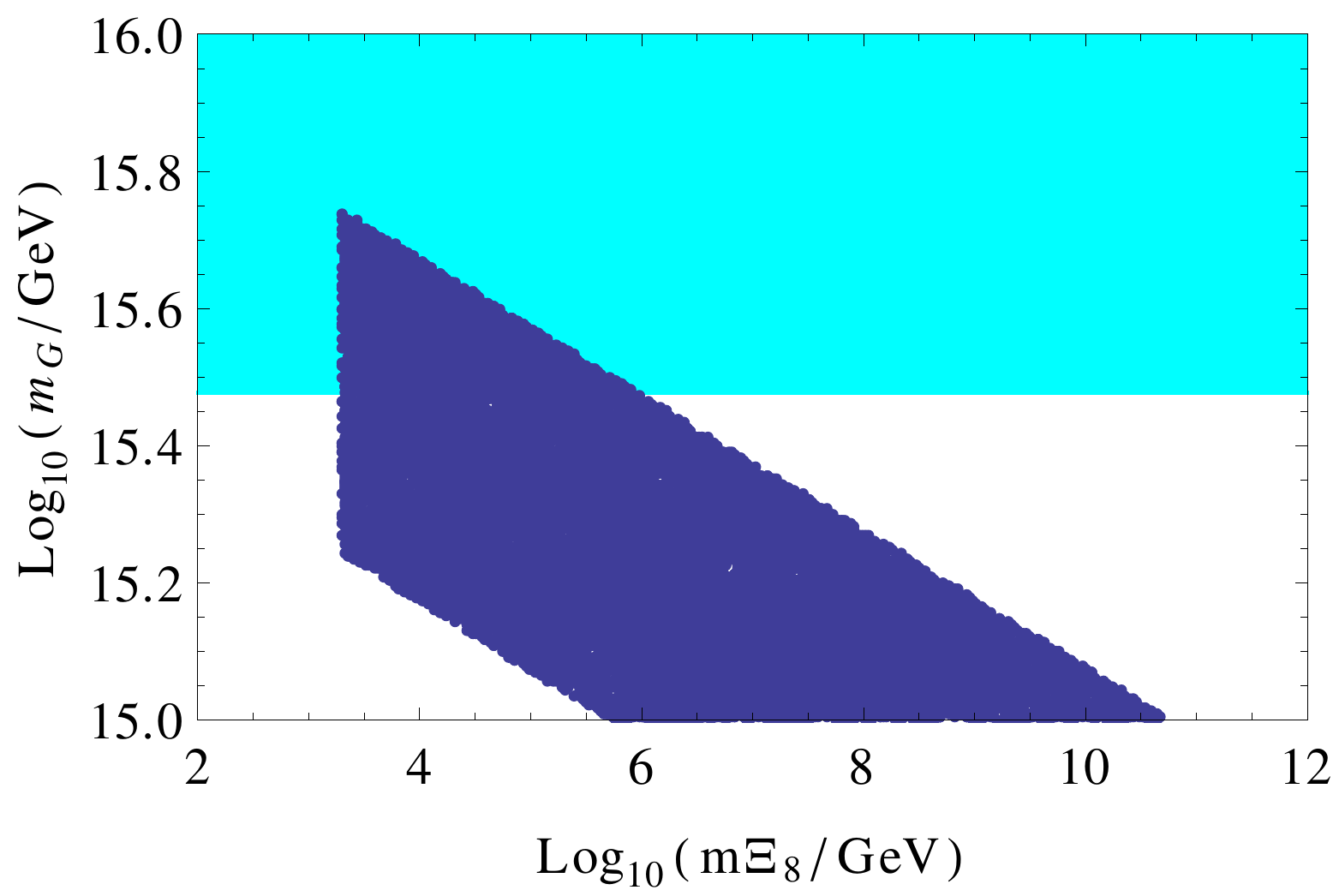}%
\end{tabular}
\includegraphics[width=0.42\textwidth]{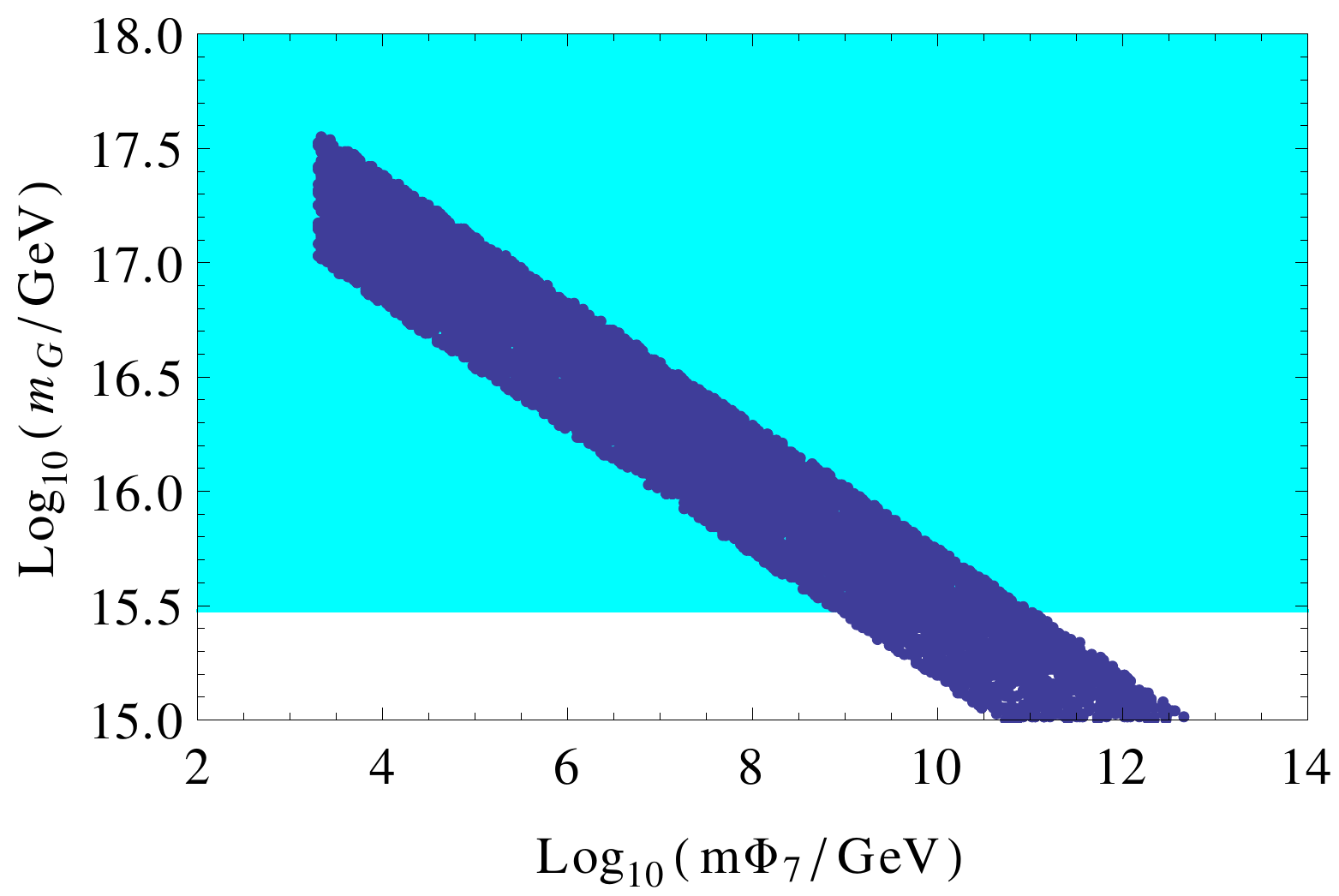} 
\end{figure}

The master equation for the running of the inverse gauge couplings at the one-loop
level is 
\begin{equation}
\alpha_{i}^{-1}(\mu)=\alpha_{i}^{-1}(\mu_{0})-\frac{b_{i}^{eff}(\mu)}{2\pi}
\ln\left(\frac{\mu}{\mu_{0}}\right),\ \ \ \ \mbox{with} \ \ \ \ i=1,2,3
\label{GUT4}
\end{equation}
The effective one-loop RGE coefficients, taking into account the thresholds
from 
particles with masses $m_{f}$, are given by 
\begin{eqnarray}  \label{thresh-1}
&&b_{i}^{eff}(\mu) = \sum_{f} \theta (\mu - m_{f})b_{i}^{f}.
\end{eqnarray}
The contribution of each particle $b^{f}_{i}$ is calculated according to 
\begin{eqnarray}  \label{Dynkin}
&&b_{i} = - \frac{11}{3} T_{i}(R_{G}) + \frac{2}{3} T_{i}(R_{F}) + \frac{1}{3%
} T_{i}(R_{B}).
\end{eqnarray}
where $T(R_{I})$ are the Dynkin indexes of the representations $R_{I}$ to
which belong $I=G$, $F$, and $B$-the gauge bosons, fermions, and scalars,
respectively. They are defined as $T(R) \delta_{mn} = \mathop{\rm Tr}%
\nolimits(T_{m}(R) T_{n}(R))$, with $T_{m}(R)$ being generators in the
representation $R$. For the lowest-dimension representations of $SU(N)$ they
are $T(fundamental) = 1/2$, $T(Adj) = N$. For the SM we have $%
(b_{3}^{SM},b_{2}^{SM},b_{1}^{SM})=(-7,-19/6,41/10)$, which correspond to
the contributions of the SM fermions and one copy of the $SU(2)_{L}$ Higgs
doublet. The additional non-SM fields introduce extra contributions $\Delta
b_{i}^{^{\prime }}$ to these coefficients %
$b_{i} = b^{SM}_{i} + \Delta b_{i}^{^{\prime }}$. In Table~\ref{tab:XI},
Appendix \ref{sec:Simple-BM}, the $\Delta b_{i}^{^{\prime }}$ contributions
of the fields in the $5_{S}$, $10_{S}$, $24_{S}$, $24_{f}$, and $45_{S}$ $%
SU(5)$ representations are shown. In order to correctly unify the gauge
couplings and fulfill all the 
requirements \textbf{(i)}-\textbf{(iv)} in Sec. \ref{sec:Setup}, which we
impose on the models, these $\Delta b^{^{\prime }}_{i}$ coefficients should
obey certain conditions. Some of the simplest configurations of the non-SM
fields with the masses below the GUT scale, which obey these conditions, are
listed in Table~\ref{tab:X}, Appendix \ref{sec:Simple-BM}. 
All the listed field configurations have a highly split mass spectrum with
type-I and type-III seesaw mediators at $m_{S} \sim 10^{14}$  GeV and remaining
scalars at 
%
$m_{NP}=2$  TeV. 
Let us note that since among the scalars there are the color octets, we use
2 TeV as a limit recently established on the mass of color octets by the CMS and
ATLAS collaborations  \cite{LHC-color-8} from the dijet pair signature
searches. We have purposely chosen the latter scale to be low enough so that the
colored scalars are within the LHC's mass reach, 
while the large value for the seesaw scale $m_{S}$ is in agreement with the
small neutrino masses. 
%
Let us consider two simplest models (1) and (2) from Table~\ref{tab:X}:\\[3mm%
]
%
\emph{Model (1)} This is the simplest of all the benchmark models passing our
conditions \textbf{(i)}-\textbf{(iv)} in Sec.  \ref{sec:Setup}, including the
unification of the gauge couplings and smallness of neutrino masses. %
However, if we fix 
the masses of the scalar fields $\Xi_{3,8}$ at $m_{NP}=2$ TeV 
and the fermionic seesaw mediators $\rho_{0,3}$, $N_{R}$ at $m_{S}=10^{14}$
GeV, as it is done for all the models in Table~\ref{tab:X}, we get the GUT
scale $m_{G}=2.7\times 10^{15}$GeV, which is in slight tension with 
the limit imposed by the proton stability $m_{G} \gtrsim 3\times 10^{15}$
GeV. This flaw can easily be cured by allowing the masses of 
the fields to vary independently in the range 
%
2 TeV$\leq m_{\Xi 3}, m_{\Xi 8} \leq m_{S}<m_{G}$. Now, as shown in Fig.~\ref%
{GUT6}, a significant part of the model parameter space corresponds to the
GUT scale in the range allowed by the proton decay constraint. From the left
panel of this figure we can see that in this part of the parameter space %
%
the masses $m_{S}$ of the seesaw mediators $\rho_{0,3}$, $N_{R}$ 
could be in the ballpark of $[10^{8.5},10^{15.5}]$ GeV. 
The right panel of Fig.~\ref{GUT6} shows that in the part of the parameter
space consistent with the proton decay constraints, the mass of the scalar
octet $\Xi_{8}$ is relatively low, 2 TeV$<m_{\Xi_{8}}\lesssim 10^6$ GeV.
Thus, there is a chance for $\Xi_{8}$ to be within the mass reach of the
current run of the LHC. However, its production cross section suffers from
several suppression factors. The process of single-$\Xi_{8}$ production is
only possible in the gluon fusion $g g\rightarrow \Xi_{8}$ via a loop with
two or three internal $\Xi_{8}$, which is suppressed by the large $%
m_{\Xi_{8}}$. On the other hand, as seen from Eq. (\ref{LY}), it cannot be
produced in $q \bar{q}\rightarrow \Xi_{8}$. The tree-level pair production
process $g g \rightarrow \Xi_{8} \Xi_{8}$, being not suppressed in the
amplitude, has a high threshold of $2 m_{\Xi_{8}}$. If produced, $\Xi_{8}$
decays in a unique channel $\Xi_{8}\rightarrow g g$. However, the
corresponding signal at the LHC could be challenging from the viewpoint of
the identification of its origin. Below we consider another benchmark model
with a more distinctive signature of the color octet from $\mathbf{45}$. 
%

\begin{figure}[htb]
\centering
\begin{tabular}{cc}
\includegraphics[width=0.5\textwidth]{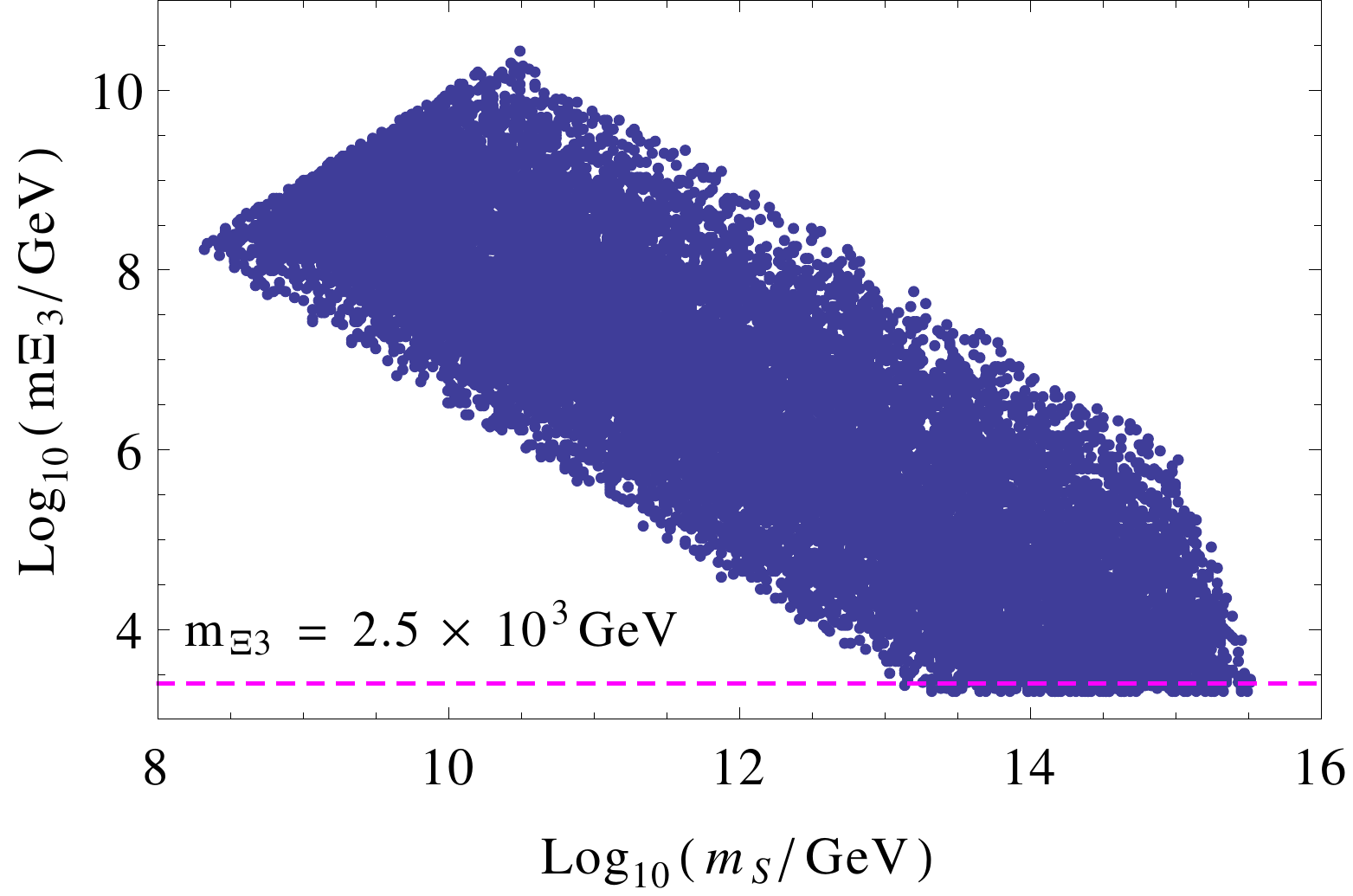} & %
\includegraphics[width=0.5\textwidth]{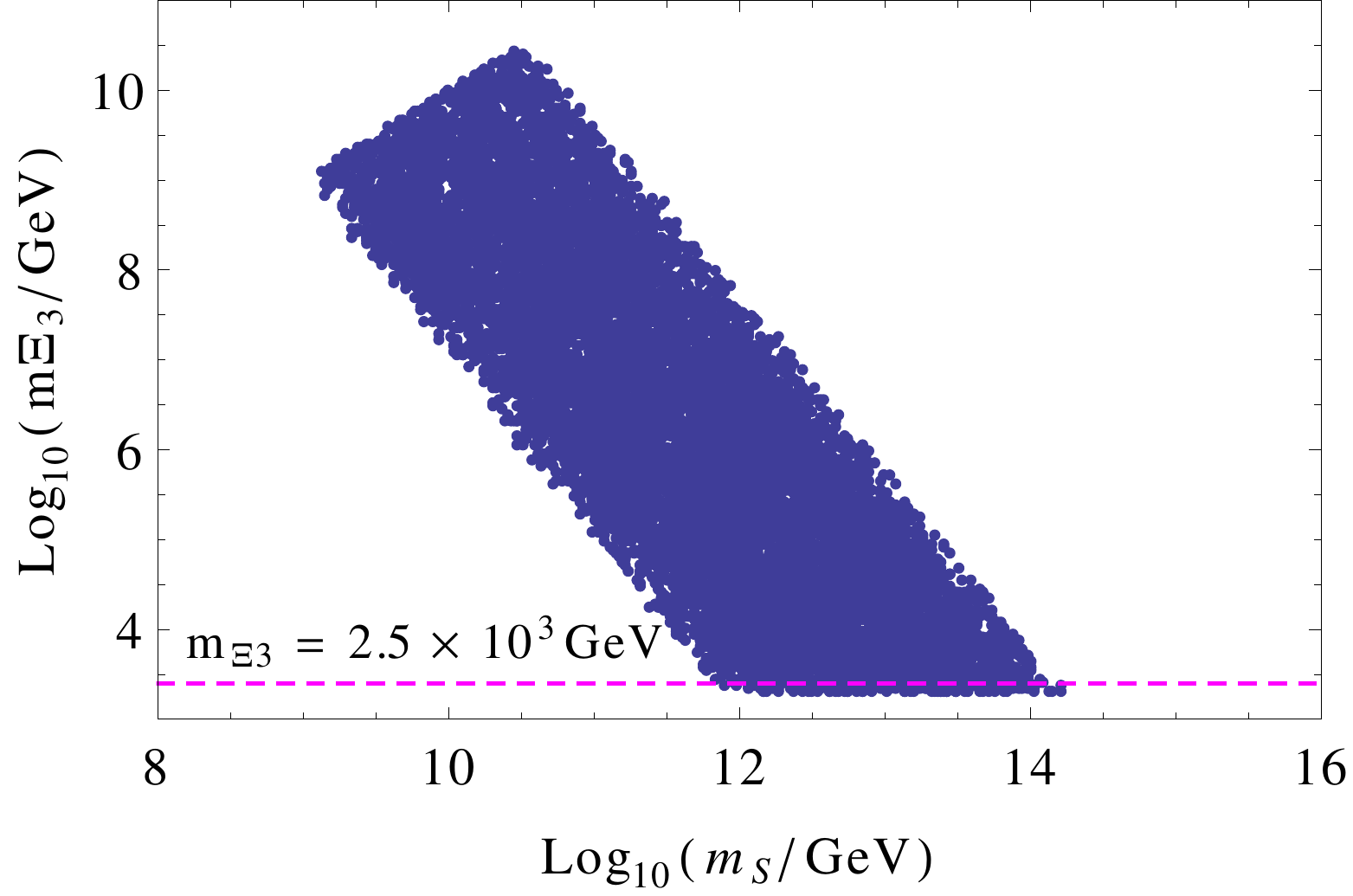}%
\end{tabular}%
\caption{Dark matter constraints: allowed parameter space in $m_{S}$ and $%
m_{\Xi 3}$. The left and right panels correspond to  \textit{Models (1)} and 
\textit{(2)}, respectively.}
\label{GUT5}
\end{figure}

A viable dark matter (DM) candidate in non-SUSY models is an issue requiring
special efforts. References \cite{Cirelli:2006,Perez:2008ry}
recently proposed the neutral component $\Xi^{0}$ of the $\Xi_{3}\sim
(1,3)_{0}\subset \mathbf{24}_{s}$ as a scalar cold dark matter (CDM)
candidate. %
%
The necessary condition for its stability is the vanishing of the trilinear
coupling of $H^{\dagger}\Xi_{3}H$ 
with the SM Higgs doublet, $H$. Also, 
its VEV 
must be zero, $\langle\Xi^{0}\rangle = 0$. These conditions, as shown in
Ref. \cite{Perez:2008ry}, can be implemented by an \emph{ad hoc} fine-tuning of the
scalar potential parameters. Unfortunately, no proper custodial symmetry,
protecting this fine-tuning, can be incorporated in the adjoint $SU(5)$
framework \cite{Perez:2008ry}. Despite this complication, we study
constraints on our model from the assumption that the $\Xi^{0}$ is a CDM
candidate. It has been shown in Ref. \cite{Cirelli:2006} that the thermal
relic abundance could be compatible with the observed DM abundance $%
\Omega_{DM}h^{2}=0.110\pm 0.005$, if $m_{\Xi_{3}}\sim 2.5$ TeV. We use this
fixed value as a condition allowing the presence of a viable CDM particle
candidate $\Xi^{0}$ in the model. In Fig.~\ref{GUT5} we show limitations on
the model parameter space, taking into account this condition. 
%
As seen, the lowest bound on the seesaw mediator mass is 
\mbox{$m_{S}
\gtrsim 10^{13}$ GeV.} %
This large value of the seesaw mass scale perfectly accounts for the
smallness of the neutrino masses. 
\\[3mm]
%
\emph{Model (2)}: This is the next simplest benchmark model from Table~\ref%
{tab:X}. It contains a scalar color octet 
\mbox{$\Phi_{7}\sim (8,2)_{1/2}
\subset {\bf 45}$.} %
%
Its production and possible signals at the LHC have been studied in 
Ref. \cite{Khalil:2014gba} (for the earlier studies of the color octet
scalars at the LHC see, for instance, Ref. \cite{Color-8-LHC-earlier}). The
neutral and charged components $S_{R,I}^{0}$, $S^{\pm}$ of this multiplet,
unlike the $\Xi_{8}$ in \textit{Model (1)}, 
can be singly produced at tree level in quark-antiquark annihilation 
$q\overline{q}\rightarrow S_{R,I}^{0}$, $q^{\prime }\overline{q}\rightarrow
S^{\pm}$ and in the gluon fusion at one-loop level via \emph{b}-quark loop, which
is much less suppressed than the loop-induced production of a single $%
\Xi_{8} $. The tree-level pair production, dominated by the gluon fusion $gg
\rightarrow S^{0}S^{0}, S^{\pm}S^{\mp}$, is approximately an order of
magnitude smaller than the single-$S$ production at $m_{S}\sim$ 2 TeV,
according to Ref. \cite{Khalil:2014gba}. However, the single-$S$ production
cannot overcome the SM background with all the versatile kinematical cuts
applied in \cite{Khalil:2014gba}, and therefore has no observational
prospects. On the other hand, it has been shown that the $S$-pair
production, even having a smaller production cross section, may get larger
than the background in the $4b$-tagged final-state jets with a cut $%
P_{T}\geq 800$ GeV. 
%

Repeating the analysis made for the previous model, we find masses $%
m_{\Phi_{7}}$ compatible with the proton decay constraints $m_{G}\geq
3\times 10^{15}$GeV. As shown in Fig.~\ref{GUT6}, we have 2 TeV$<m_{\Phi 7}
\lesssim 10^{11}$ GeV. 
%
%
%

Let us note also that, even if we fix the mass of all the scalars at $2$TeV
and the seesaw mediators at $10^{14}$ GeV, as is done for the models in
Table~\ref{tab:X}, the GUT scale is still high enough to be safe, facing new
possible improvement of the proton decay constraints in the future
experiments. 
Now, as with \textit{Model (1)}, we consider the constraints imposed
on the present model by the interpretation of the $\Xi_{3}$ as a CDM
candidate. As we discussed above, this requires that $m_{\Xi_{3}}\sim 2.5$
TeV. Scanning the parameter space, as with \textit{Model (1)}, we
find the plot shown in the right panel of Fig.~\ref{GUT5}, from which we
derive the lower bound $m_{S} \gtrsim 10^{12}$ GeV. 
%

\section{Conclusions and Discussion}

\label{Summary}

We proposed a version of the adjoint $SU(5)$ grand unification model with an
extra $T_{7}\otimes Z_{2}\otimes Z_{3}\otimes Z_{4}\otimes Z_{4}^{\prime
}\otimes Z_{12}$ flavor symmetry, which successfully describes the SM
fermion mass and mixing pattern. The model has in total 16 effective free
parameters, from which 2 are fixed and 14 are fitted to reproduce the
experimental values of 18 observables in the quark and lepton sectors i.e.,
9 charged fermion masses, 2 neutrino mass-squared splittings, 3 lepton
mixing parameters, 3 quark mixing angles, and 1 CP-violating phase of the CKM
quark mixing matrix. 
The observed hierarchy of charged fermion masses and quark mixing angles is
a consequence of the $Z_{3}\otimes Z_{4}\otimes Z_{12}$ symmetry breaking,
triggered by the $SU(5)$ scalar singlets $\sigma $, $\tau $, and $\varphi $,
charged under this symmetry, and which acquire VEVs at very high-energy
scale, close to the GUT scale. The non-SM fermion spectrum of our model is
composed of a single heavy $SU(5)$ singlet right-handed Majorana neutrino $%
N_{R}$ and two more fermionic fields: an electroweak singlet $\rho_{0}$ and
triplet $\rho_{3}$, both from the adjoint $\mathbf{24}$ irrep of $SU(5)$.
Thus the light neutrino masses arise in our model from type-I and type-III
seesaw mechanisms mediated by these fields. 
The smallness of neutrino masses is a consequence of their inverse scaling
with respect to the masses of these three seesaw mediators as well as the
quadratic proportionality to the presumably small neutrino Yukawa couplings.
The model predictions for the physical parameters in the quark and lepton
sectors are in  good agreement 
with the experimental data. The experimentally observed deviation from the
trimaximal pattern is implemented by introducing two $T_{7}$ triplet scalars 
$\chi$ and $\xi$, singlets under $SU(5)$. The model predicts an effective
Majorana neutrino mass, relevant for neutrinoless double beta decay, with
values $m_{\beta \beta }=$ 4 and 50 meV, for the normal and the inverted
neutrino spectrum, respectively. In the latter case, our prediction is within
the declared reach of the next-generation bolometric CUORE experiment \cite%
{Alessandria:2011rc} or, more realistically, of the next-to-next-generation
tone-scale $0\nu\beta\beta$-decay experiments.

According to our model, the 
leptonic Dirac CP-violating phase is vanishing. 
In view of the current experimental trend, it could get into tension with the
observations. %
%
The latest fit of the neutrino oscillation experimental data by the Valencia
group, Ref. \cite{Forero:2014bxa}, indicates a nonvanishing CP-violating
phase at the $1\sigma $ (less than $2\sigma $) level for the normal (inverted) neutrino mass hierarchy. 
The Bali group in their fits claims slightly larger significance, but also
below the $3\sigma $ level \cite{Capozzi:2013csa}. Thus, with the current
experimental significance, essentially below than the golden $5\sigma$, our
model is not yet in trouble with the CP. On the other hand, we are aware of
the necessity of reconsidering this aspect of our approach, which will be
done elsewhere.

%

In the last section of this paper, we studied the compatibility of our model
with certain physical conditions expected from a plausible GUT model. Among
them we considered the gauge coupling unification, the proton stability
constraints, the smallness of the active neutrino masses, and nontrivial LHC
phenomenology. Towards this end, we specified the simplest benchmark models
of the non-SM particle spectrum lighter than the GUT scale and which meet
these conditions. We examined two of them and found that they may give rise
to observable signals at the LHC, as well as contain a viable scalar CDM
particle candidate. In order to account for the observed DM relic abundance
in the Universe, the latter imposes certain constraints on the model
parameter space. %
%
From this DM condition we found, in particular, the lower limits in the mass
of 
the seesaw type-III mediator ($m_{S}$), for the two simple benchmark models
analyzed in the paper. 
%
It is worth noting that  \textit{Model (1)} is manifestly 
falsifiable, as seen from Fig.\ref{GUT6}. 
A relatively small improvement of the proton decay lifetime lower limit up
to $\sim 6\times 10^{15}$ GeV would reject the model. 

%

Various aspects of the adjoint $SU(5)$ scenario have been studied in the
literature 
%
\cite{Perez:2007rm,Perez:2008ry,Blanchet:2008cj}. In Refs. \cite{Perez:2007rm,Perez:2008ry}, the fact that $\rho_{3}$ is the lightest field
in the $\mathbf{24}$ 
%
offers a viable way to understand the baryogenesis via leptogenesis. 
In this work, we also considered the field $\rho_{3}$ as the lightest
component of the $\mathbf{24}$, and therefore, the baryogenesis can be
accomplished in our model in the same way. 
Note also that in this mechanism \cite{Blanchet:2008cj}, the decays of the $%
\rho_{3}$ create a lepton asymmetry, which then 
is converted in a baryon asymmetry by the sphalerons. Imposing the condition
of the successful leptogenesis and for the normal hierarchy of neutrinos,
one finds that the mass of $\rho_{3}$ should be large, which is in agreement
with the large $\rho_{3}$ mass values predicted in the present paper.\newline
Finally, as was previously noticed in Ref. \cite{Perez:2008ry}, the mass
of $\rho_{8}$ must be heavier than $10^{6}$ GeV$-10^{7}$ GeV in order to
satisfy the constraints from the big bang nucleosynthesis for the GUT scales
larger than $3\times 10^{15}$ GeV. This condition is also consistent with
our results, where the mass of this field is set around the GUT scale. 

\section*{Acknowledgments}

We are grateful to Martin Hirsch for valuable discussions and
recommendations. We also thank Ernest Ma and Ivan Girardi for useful
comments. This work was partially supported by Fondecyt (Chile), Grants
No.~11130115, No.~1150792, No.~1140390, No.~3150472, and by DGIP internal
Grant No.~111458. 

\appendix

\section{The product rules for $T_{7}$ \label{A}}

\label{A} The group $T_{7}$ is the minimal non-Abelian discrete group having
a complex triplet. The discrete group $T_{7}$ has $21$ elements and $5$
irreps, i.e., one triplet $\mathbf{3}$, one antitriplet $\mathbf{\bar{3}}$,
and three singlets $\mathbf{1}_{0}$, $\mathbf{1}_{1}$ and $\mathbf{1}_{2}$ 
\cite{Ishimori:2010au}. Furthermore, the $T_{7}$ group is a subgroup of $%
SU(3)$ and $\Delta (3N^{2})$ with $N=7$ and is isomorphic to $Z_{7}\rtimes
Z_{3}$. The triplet and antitriplet irreducible representations can be
defined as follows \cite{Ishimori:2010au}: 
\begin{equation}
\mathbf{3}\equiv \3tvec{x_{1}}{x_{2}}{x_{4}},\quad \mathbf{\bar{3}}\equiv \3%
tvec{x_{-1}}{x_{-2}}{x_{-4}}=\3tvec{x_{6}}{x_{5}}{x_{3}}.
\end{equation}%
The triplet and antitriplet $T_{7}$ tensor irreducible representations
satisfy the following product rules: 
\begin{eqnarray}
\3tvec{x_{1}}{x_{2}}{x_{4}}_{\mathbf{3}}\otimes \3tvec{y_{1}}{y_{2}}{y_{4}}_{%
\mathbf{3}} &=&\3tvec{x_{2}y_{4}}{x_{4}y_{1}}{x_{1}y_{2}}_{\mathbf{\bar{3}}%
}\oplus \3tvec{x_{4}y_{2}}{x_{1}y_{4}}{x_{2}y_{1}}_{\mathbf{\bar{3}}}\oplus %
\3tvec{x_{4}y_{4}}{x_{1}y_{1}}{x_{2}y_{2}}_{\mathbf{3}}, \\
\3tvec{x_{6}}{x_{5}}{x_{3}}_{\mathbf{\bar{3}}}\otimes \3tvec{y_{6}}{y_{5}}{%
y_{3}}_{\mathbf{\bar{3}}} &=&\3tvec{x_{5}y_{3}}{x_{3}y_{6}}{x_{6}y_{5}}_{%
\mathbf{3}}\oplus \3tvec{x_{3}y_{5}}{x_{6}y_{3}}{x_{5}y_{6}}_{\mathbf{3}%
}\oplus \3tvec{x_{3}y_{3}}{x_{6}y_{6}}{x_{5}y_{5}}_{\mathbf{\bar{3}}}, \\
\3tvec{x_{1}}{x_{2}}{x_{4}}_{\mathbf{3}}\otimes \3tvec{y_{6}}{y_{5}}{y_{3}}_{%
\mathbf{\bar{3}}} &=&\3tvec{x_{2}y_{6}}{x_{4}y_{5}}{x_{1}y_{3}}_{\mathbf{3}%
}\oplus \3tvec{x_{1}y_{5}}{x_{2}y_{3}}{x_{4}y_{6}}_{\mathbf{\bar{3}}}  \notag
\\
&&\oplus \sum_{k=0,1,2}(x_{1}y_{6}+\omega ^{k}x_{2}y_{5}+\omega
^{2k}x_{4}y_{3})_{\mathbf{1}_{k}}.
\end{eqnarray}%
%
%
%
%
%
%
%
%
%
%
%
%
%
%
%
%
%
%
Whereas the tensor products between singlets are given by the relations 
\begin{eqnarray}
(x)_{\mathbf{1}_{0}}(y)_{\mathbf{1}_{0}} &=&(x)_{\mathbf{1}_{1}}(y)_{\mathbf{%
1}_{2}}=(x)_{\mathbf{1}_{2}}(y)_{\mathbf{1}_{1}}=(xy)_{\mathbf{1}_{0}},~ 
\notag \\
(x)_{\mathbf{1}_{1}}(y)_{\mathbf{1}_{1}} &=&(xy)_{\mathbf{1}_{2}},~  \notag
\\
(x)_{\mathbf{1}_{2}}(y)_{\mathbf{1}_{2}} &=&(xy)_{\mathbf{1}_{1}},
\end{eqnarray}%
%
%
%
%
%
%
%
%
%
%
%
%
%
%
%
%
%
%
%
the product rules between triplets and singlets read 
\begin{equation}
(y)_{\mathbf{1}_{k}}\otimes \3tvec{x_{1(6)}}{x_{2(5)}}{x_{4(3)}}_{\mathbf{3(%
\bar{3})}}=\3tvec{yx_{1(6)}}{yx_{2(5)}}{yx_{4(3)}}_{\mathbf{3(\bar{3})}}.
\end{equation}%
where $\omega =e^{i\frac{2\pi }{3}}$. The representation $\mathbf{1}_{0}$ is
trivial, while the nontrivial $\mathbf{1}_{1}$ and $\mathbf{1}_{2}$ are
complex conjugate to each other. Some reviews of discrete symmetries in
particle physics can be found in Refs. \cite%
{King:2013eh,Altarelli:2010gt,Ishimori:2010au,Discret-Group-Review}.


\section{On the universality of Yukawa couplings}

\label{sec:YC-univ} The aforementioned scheme of approximate universality of
dimensionless couplings can be justified by adding an extra $Z_{24}$
symmetry and four $SU(5)$ scalar singlets, assigned as $T_{7}$ trivial
singlets, as well as by setting their VEVs to be equal to $\lambda \Lambda $%
, with $\lambda =0.225$, one of the Wolfenstein parameters, and $\Lambda $
being our model cutoff. One of the four $SU(5)$ scalar singlets, namely $S$,
can be assumed to have the same $Z_{12}$ charge as the scalar field $\sigma $
of our model and can also be made charged under the new $Z_{24}$ symmetry
and neutral under the remaining cyclic symmetries. The remaining three $SU(5)
$ scalar singlets, namely $\Delta _{i}$ $(i=1,2,3)$, can be assumed to be
only charged under the new $Z_{24}$ symmetry. These four new scalar fields
will transform under the $Z_{24}$ symmetry as follows: 
\begin{equation}
S\rightarrow e^{-i\frac{\pi }{6}}S,\quad \Delta _{1}\rightarrow i\Delta
_{1},\quad \Delta _{2}\rightarrow e^{i\frac{\pi }{3}}\Delta _{2},\quad
\Delta _{3}\rightarrow e^{i\frac{\pi }{4}}\Delta _{3}.
\end{equation}%
The aforementioned $Z_{24}$ charge assignments will generate the following $%
Z_{24}$ neutral combinations of the new scalar fields: 
\begin{equation}
S^{6}\Delta _{1}^{2},S^{4}\Delta _{2}^{2},\quad S^{3}\Delta _{3}^{2}.
\end{equation}%
These $Z_{24}$ neutral combinations will give rise to the following Yukawa
operators invariant under the symmetries of the model: 
\begin{equation}
\varepsilon ^{ijklp}\Psi _{ij}^{\left( 1\right) }H_{p}^{\left( 2\right)
}\Psi _{kl}^{\left( 1\right) }\frac{S^{6}\Delta _{1}^{2}}{\Lambda ^{8}}%
,\quad \varepsilon ^{ijklp}\Psi _{ij}^{\left( 2\right) }H_{p}^{\left(
3\right) }\Psi _{kl}^{\left( 2\right) }\frac{S^{4}\Delta _{2}^{2}}{\Lambda
^{6}},\quad \varepsilon ^{ijklp}\Psi _{ij}^{\left( 1\right) }H_{p}^{\left(
3\right) }\Psi _{kl}^{\left( 3\right) }\frac{S^{3}\Delta _{3}^{2}}{\Lambda
^{5}},\quad \varepsilon ^{ijklp}\Psi _{ij}^{\left( 3\right) }H_{p}^{\left(
3\right) }\Psi _{kl}^{\left( 1\right) }\frac{S^{3}\Delta _{3}^{2}}{\Lambda
^{5}}.
\end{equation}%
The first two aforementioned Yukawa operators will contribute to the $11$
and $22$ entries of the up-type quark mass matrix, whereas the last two
operators will contribute to the $13$ and $31$ entries. These new
contributions will be proportional to $\lambda ^{8}$, $\lambda ^{6}$, and $\lambda ^{5}$, respectively, and thus will introduce deviations from the
exact universality in the dimensionless Yukawa couplings. These
aforementioned contributions will correspond to the first-order term in the $ \lambda $ expansion of the expressions given in Eq. (\ref{apuniversality}).

\section{Simple benchmark models}

\label{sec:Simple-BM}

\noindent The contributions to the $\Delta b^{^{\prime }}$ coefficients for
each field in the $5_{s}$, $24_{s}$, $45_{s}$, and $24_{f}$ reps are shown in
Table~\ref{tab:XI}. In Table~\ref{tab:X}, the configurations which generate
neutrino mass in agreement with the mechanism described in Sec. III are
shown. Each particle content, added to the SM, lead to \emph{"SM+X"}
configurations unifying gauge couplings almost equal to or better than the
MSSM, i.e, each one of the simplest configurations 
satisfies: $\alpha_{2}^{-1}(m_{G})-\alpha_{1}^{-1}(m_{G})\lesssim \alpha_{2
MSSM}^{-1}(m_{G})-\alpha_{1 MSSM}^{-1}(m_{G}) $ and $3\times 10^{15}$ GeV $%
\lesssim m_{G}\leq 10^{18}$ GeV in order to obtain proton lifetimes allowed
by the actual experimental bounds. 
%
\begin{table}[h]
\begin{center}
\scalebox{0.85}{
\begin{tabular}{| l | l |l |}
\hline
 $SU(5)$ Rep & Fields & $\Delta b'_{1,2,3}$  \\ \hline

 $5_{S}$ & $H_{1}, H_{2}$ & $(\frac{1}{10},\frac{1}{6},0), (\frac{1}{15},0,\frac{1}{6})$  \\ \hline
                
 $24_{S}$ & $\Xi_{1}$, $\Xi_{3}$, $\Xi_{8}$, $\Xi_{(3,2)}$, $\Xi_{(\overline{3},2)}$ & $(0,0,0)$, $(0,\frac{2}{3},0)$,  $(0,0,1)$, $(\frac{5}{6},\frac{1}{2},\frac{1}{3})$, $(\frac{5}{6},\frac{1}{2},\frac{1}{3})$  \\ \hline       
                                          
 $45_{S}$ & $\Phi_{1}$, $\Phi_{2}$, $\Phi_{3}$, $\Phi_{4}$, $\Phi_{5}$, $\Phi_{6}$, $\Phi_{7}$ & $(\frac{1}{10},\frac{1}{6},0)$, $(\frac{1}{15},0,\frac{1}{6})$ ,  $(\frac{1}{5},2,\frac{1}{2})$  $(\frac{4}{15},0,\frac{1}{6})$, $(\frac{1}{30},\frac{1}{2},\frac{1}{3})$  $(\frac{2}{15},0,\frac{5}{6})$ $(\frac{4}{5},\frac{4}{3},2)$\\ \hline

$24_{f}$ & $\rho_{1}$, $\rho_{3}$, $\rho_{8}$, $\rho_{(3,2)}$, $\rho_{(\overline{3},2)}$ & $(0,0,0)$, $(0,\frac{4}{3},0)$,  $(0,0,2)$, $(\frac{5}{3},1,\frac{2}{3})$, $(\frac{5}{3},1,\frac{2}{3})$  \\ \hline   
                                          
\end{tabular}} 
\end{center}
\caption{ Contribution to the running of the gauge couplings of the fields
from the $5_{s}$, $24_{s}$, $45_{s}$ and $24_{f}$ representations of the $%
SU(5)$. The seesaw particles mediators $N_{R}$ and $\protect\rho_{0,3}$ are
added at the $10^{14}$ GeV, contributing to the running of the gauge
couplings with $(0,4/3,0)$ in the range $[10^{14} GeV, m_{G}]$. The
contribution of the fields in the $24_{f}$ is twice the contribution of $%
24_{s}$ owing its fermionic nature. }
\label{tab:XI}
\end{table}

\begin{table}[hh]
\begin{center}
\scalebox{0.95}{
\begin{tabular}{|c| l | l | l |  l |}
\hline
(X)& Model (X) & $\Delta b'_{1,2,3}$ & $m_{G}$ & $SU(5)$ Reps \\ \hline

(1)&  $\Xi_{8}+2\Xi_{3}+N_{R}+\rho_{0}+\rho_{3}$ & $(0,\frac{4}{3},1)$ & $2.82 \times 10^{15}$ & $1+24_{s}+24_{f}$ \\ \hline
   
(2)&  $2\Phi_{7}+\Phi_{1}+\Xi_{3}+N_{R}+\rho_{0}+\rho_{3}$ & $(\frac{17}{10},\frac{7}{2},4)$ & $4.17 \times 10^{17}$ & $1+24_{s}+45_{s}+24_{f}$ \\ \hline

(3)&  $2\Phi_{7}+2\Phi_{1}+\Xi_{3}+N_{R}+\rho_{0}+\rho_{3}$ & $(\frac{9}{5},\frac{11}{3},4)$ & $1.34 \times 10^{17}$ & $1+24_{s}+45_{s}+24_{f}$ \\ \hline
             
(4)&  $2\Xi_{3}+\Phi_{6}+\Phi_{7}+N_{R}+\rho_{0}+\rho_{3}$ & $(\frac{14}{15},\frac{8}{3},\frac{17}{6})$ & $4.74 \times 10^{16}$ & $1+24_{s}+45_{s}+24_{f}$ \\ \hline       

 (5)& $3\Xi_{3}+2\Xi_{8}+\Xi_{(3,2)}+N_{R}+\rho_{0}+\rho_{3}$ & $(\frac{5}{6},\frac{5}{2},\frac{7}{3})$ & $6.8 \times 10^{15}$ & $1+24_{s}+45_{s}+24_{f}$ \\ \hline

 (6)& $\Phi_{2}+2\Xi_{3}+\Xi_{8}+N_{R}+\rho_{0}+\rho_{3}$ & $(\frac{1}{15},\frac{4}{3},\frac{7}{6})$ & $6.8 \times 10^{15}$ & $1+24_{s}+45_{s}+24_{f}$ \\ \hline
                                           
(7)& $\Phi_{2}+3\Xi_{3}+2\Xi_{8}+\Xi_{(3,2)}+N_{R}+\rho_{0}+\rho_{3}$ & $(\frac{9}{10},\frac{5}{2},\frac{5}{2})$ & $1.7 \times 10^{16}$ & $1+24_{s}+45_{s}+24_{f}$ \\ \hline

 (8)& $2\Phi_{6}+\Phi_{3}+N_{R}+\rho_{0}+\rho_{3}$ & $(\frac{7}{15},2,\frac{13}{6})$ & $4.6 \times 10^{16}$ & $1+45_{s}+24_{f}$ \\ \hline
                      
(9)&   $\Phi_{2}+2\Phi_{7}+\Phi_{3}+\Phi_{4}+N_{R}+\rho_{0}+\rho_{3}$ & $(\frac{44}{15},\frac{14}{3},\frac{29}{6})$ & $4.63 \times 10^{16}$ & $1+45_{s}+24_{f}$ \\ \hline

 (10)& $2\Phi_{7}+\Phi_{3}+\Phi_{4}+\Xi_{(3,2)}+N_{R}+\rho_{0}+\rho_{3}$ & $(\frac{37}{10},\frac{31}{6},5)$ & $6.77 \times 10^{15}$ & $1+24_{s}+45_{s}+24_{f}$ \\ \hline

 (11)& $3\Phi_{7}+\Phi_{3}+\Phi_{4}+\Xi_{(3,2)}+N_{R}+\rho_{0}+\rho_{3}$ & $(\frac{9}{2},\frac{13}{2},7)$ & $4.17 \times 10^{17}$ & $1+24_{s}+45_{s}+24_{f}$ \\ \hline
              
 (12)& $\Phi_{6}+2\Phi_{7}+\Phi_{3}+3\Xi_{(3,2)}+N_{R}+\rho_{0}+\rho_{3}$ & $(\frac{133}{30},\frac{37}{6},\frac{19}{3})$ & $4.63 \times 10^{16}$ & $1+24_{s}+45_{s}+24_{f}$ \\ \hline
                                          
\end{tabular}} 
\end{center}
\caption{Simple X configurations, leading to the GCU and generate the
neutrino mass via type-I and type-III seesaw mechanisms. The masses of all
the extra scalar particles $\Phi_{i}$ and $\Xi_{i}$ are about $%
m_{NP}=2\times 10^{3} GeV$ while the the fermionic seesaw mediators $N_{R}, 
\protect\rho_{0,3}$, are much heavier with the masses at the scale $%
m_{S}=10^{14}$ GeV. The $\Delta b^{^{\prime }}$s RGE coefficients correspond
to the contribution of the scalars. The seesaw mediators contribute with an
extra $(0,4/3,0)$ above the scale $10^{14} GeV$.}
\label{tab:X}
\end{table}


\begin{thebibliography}{999}
\bibitem{LHC-H-discovery} G. Aad et al. [ATLAS Collaboration], Phys. Lett. B 
\textbf{716}, 1 (2012); S. Chatrchyan et al. [CMS Collaboration], Phys.
Lett. B \textbf{716}, 30 (2012).


\bibitem{PDG} J. Beringer et al. (Particle Data Group), Phys.\ Rev.\ D 
\textbf{86} 010001 (2012).

\bibitem{An:2012eh} F.~P.~An \textit{et al.} [DAYA-BAY Collaboration], 
Phys.\ Rev.\ Lett.\ \textbf{108}, 171803 (2012) [arXiv:1203.1669 [hep-ex]].

\bibitem{Abe:2011sj} K.~Abe \textit{et al.} [T2K Collaboration], 
Phys.\ Rev.\ Lett.\ \textbf{107}, 041801 (2011) [arXiv:1106.2822 [hep-ex]].

\bibitem{Adamson:2011qu} P.~Adamson \textit{et al.} [MINOS Collaboration], 
Phys.\ Rev.\ Lett.\ \textbf{107}, 181802 (2011) [arXiv:1108.0015 [hep-ex]].

\bibitem{Abe:2011fz} Y.~Abe \textit{et al.} [DOUBLE-CHOOZ Collaboration], 
Phys.\ Rev.\ Lett.\ \textbf{108}, 131801 (2012) [arXiv:1112.6353 [hep-ex]].

\bibitem{Ahn:2012nd} J.~K.~Ahn \textit{et al.} [RENO Collaboration], 
Phys.\ Rev.\ Lett.\ \textbf{108}, 191802 (2012) [arXiv:1204.0626 [hep-ex]].


\bibitem{Forero:2014bxa} 
D.~V.~Forero, M.~Tortola and J.~W.~F.~Valle, 
Phys.\ Rev.\ D \textbf{90}, no. 9, 093006 (2014) [arXiv:1405.7540 [hep-ph]]. 

\bibitem{Fritzsch:1999ee} H.~Fritzsch and Z.~-z.~Xing, 
Prog.\ Part.\ Nucl.\ Phys.\ \textbf{45}, 1 (2000) [hep-ph/9912358].

\bibitem{Altarelli:2002hx} G.~Altarelli and F.~Feruglio, 
Springer Tracts Mod.\ Phys.\ \textbf{190}, 169 (2003) [hep-ph/0206077].

\bibitem{Altarelli:2010gt} G.~Altarelli and F.~Feruglio, 
Rev.\ Mod.\ Phys.\ \textbf{82} 2701 (2010) [arXiv:1002.0211 [hep-ph]]. 


\bibitem{Ishimori:2010au} H.~Ishimori, T.~Kobayashi, H.~Ohki, Y.~Shimizu,
H.~Okada and M.~Tanimoto, 
Prog.\ Theor.\ Phys.\ Suppl.\ \textbf{183}, 1 (2010) [arXiv:1003.3552
[hep-th]]. 

\bibitem{FlavorSymmRev} 
S.~F.~King, A.~Merle, S.~Morisi, Y.~Shimizu and M.~Tanimoto, 
New J.\ Phys.\ \textbf{16}, 045018 (2014) [arXiv:1402.4271 [hep-ph]].

\bibitem{Textures} M.~Gupta and G.~Ahuja, 
Int.\ J.\ Mod.\ Phys.\ A \textbf{27}, 1230033 (2012) [arXiv:1302.4823
[hep-ph]]; 
H.~Pas and E.~Schumacher, 
Phys.\ Rev.\ D \textbf{89}, no. 9, 096010 (2014) [arXiv:1401.2328 [hep-ph]]; 
A.~E.~C\'arcamo~Hern\'andez, E.~C.~Mur and R.~Martinez, 
Phys.\ Rev.\ D \textbf{90}, no. 7, 073001 (2014) [arXiv:1407.5217 [hep-ph]]. 
A.~E.~C\'arcamo~Hern\'andez and I.~d.~M.~Varzielas, 
J.\ Phys.\ G \textbf{42}, no. 6, 065002 (2015) [arXiv:1410.2481 [hep-ph]]; 
H.~Nishiura and T.~Fukuyama, 
Mod.\ Phys.\ Lett.\ A \textbf{29}, 1450147 (2014) [arXiv:1405.2416 [hep-ph]]; 
M.~Frank, C.~Hamzaoui, N.~Pourtolami and M.~Toharia, 
Phys.\ Lett.\ B \textbf{742}, 178 (2015) [arXiv:1406.2331 [hep-ph]]; 
R.~Sinha, R.~Samanta and A.~Ghosal, 
arXiv:1508.05227 [hep-ph]; 
H.~Nishiura and T.~Fukuyama, 
arXiv:1510.01035 [hep-ph]; 
R.~R.~Gautam, M.~Singh and M.~Gupta, 
Phys.\ Rev.\ D \textbf{92}, no. 1, 013006 (2015) [arXiv:1506.04868
[hep-ph]]; 
H.~Pas and E.~Schumacher, 
arXiv:1510.08757 [hep-ph]. 


\bibitem{Marzocca:2011dh} D.~Marzocca, S.~T.~Petcov, A.~Romanino and
M.~Spinrath, 
JHEP \textbf{1111}, 009 (2011) [arXiv:1108.0614 [hep-ph]].

\bibitem{Antusch:2013kna} S.~Antusch, C.~Gross, V.~Maurer and C.~Sluka, 
Nucl.\ Phys.\ B \textbf{877}, 772 (2013) [arXiv:1305.6612 [hep-ph]]. 

\bibitem{Fichet:2014vha} S.~Fichet, B.~Herrmann and Y.~Stoll, 
Phys.\ Lett.\ B \textbf{742}, 69 (2015) [arXiv:1403.3397 [hep-ph]].

\bibitem{Chen:2013wba} M.~-C.~Chen, J.~Huang, K.~T.~Mahanthappa and
A.~M.~Wijangco, 
JHEP \textbf{1310}, 112 (2013) [arXiv:1307.7711 [hep-ph]].

\bibitem{King:2012in} S.~F.~King, C.~Luhn and A.~J.~Stuart, 
Nucl.\ Phys.\ B \textbf{867}, 203 (2013) [arXiv:1207.5741 [hep-ph]].

\bibitem{Meloni:2011fx} D.~Meloni, 
JHEP \textbf{1110}, 010 (2011) [arXiv:1107.0221 [hep-ph]].

\bibitem{BhupalDev:2012nm} P.~S.~Bhupal Dev, B.~Dutta, R.~N.~Mohapatra and
M.~Severson, 
Phys.\ Rev.\ D \textbf{86}, 035002 (2012) [arXiv:1202.4012 [hep-ph]].

\bibitem{Babu:2009nn} K.~S.~Babu and Y.~Meng, 
Phys.\ Rev.\ D \textbf{80}, 075003 (2009) [arXiv:0907.4231 [hep-ph]].

\bibitem{Babu:2011mv} K.~S.~Babu, K.~Kawashima and J.~Kubo, 
Phys.\ Rev.\ D \textbf{83}, 095008 (2011) [arXiv:1103.1664 [hep-ph]].

\bibitem{Gomez-Izquierdo:2013uaa} J.~C.~G\'{o}mez-Izquierdo,
F.~G.~'al.~Canales and M.~Mondrag\'{o}n, 
arXiv:1312.7385 [hep-ph].


\bibitem{Antusch:2010es} S.~Antusch, S.~F.~King and M.~Spinrath, 
Phys.\ Rev.\ D \textbf{83}, 013005 (2011) [arXiv:1005.0708 [hep-ph]]. 

\bibitem{Hagedorn:2010th} C.~Hagedorn, S.~F.~King and C.~Luhn, 
JHEP \textbf{1006}, 048 (2010) [arXiv:1003.4249 [hep-ph]].

\bibitem{Ishimori:2008fi} H.~Ishimori, Y.~Shimizu and M.~Tanimoto, 
Prog.\ Theor.\ Phys.\ \textbf{121}, 769 (2009) [arXiv:0812.5031 [hep-ph]].

\bibitem{Patel:2010hr} K.~M.~Patel, 
Phys.\ Lett.\ B \textbf{695}, 225 (2011)

\bibitem{Cooper:2010ik} I.~K.~Cooper, S.~F.~King and C.~Luhn, 
Phys.\ Lett.\ B \textbf{690}, 396 (2010) [arXiv:1004.3243 [hep-ph]]. 

\bibitem{Emmanuel-Costa:2013gia} D.~Emmanuel-Costa, C.~Simoes and
M.~Tortola, 
JHEP \textbf{1310}, 054 (2013) [arXiv:1303.5699 [hep-ph]].

\bibitem{Chen:2007} M.~-C.~Chen and K.~T.~Mahanthappa, 
Phys.\ Lett.\ B \textbf{652}, 34 (2007) [arXiv:0705.0714 [hep-ph]].


\bibitem{Antusch:2014poa} S.~Antusch, I.~de Medeiros Varzielas, V.~Maurer,
C.~Sluka and M.~Spinrath, 
JHEP \textbf{1409}, 141 (2014) [arXiv:1405.6962 [hep-ph]].

\bibitem{Gehrlein:2014wda} J.~Gehrlein, J.~P.~Oppermann, D.~Schafer and
M.~Spinrath, 
Nucl.\ Phys.\ B \textbf{890}, 539 (2014) [arXiv:1410.2057 [hep-ph]].


\bibitem{Campos:2014lla} M.~D.~Campos, A.~E.~C\'arcamo Hern\'andez,
S.~Kovalenko, I.~Schmidt and E.~Schumacher, 
Phys.\ Rev.\ D \textbf{90}, 016006 (2014) [arXiv:1403.2525 [hep-ph]].

\bibitem{Carcamo:2014} A.~E.~C\'{a}rcamo Hern\'{a}ndez, S.~G.~Kovalenko and
I.~Schmidt, arXiv:1411.2913 [hep-ph].

\bibitem{Gehrlein:2015dxa} J.~Gehrlein, S.~T.~Petcov, M.~Spinrath and
X.~Zhang, 
arXiv:1502.00110 [hep-ph].

\bibitem{Bjorkeroth:2015ora} F.~Bjorkeroth, F.~J.~de Anda,
I.~d.~M.~Varzielas and S.~F.~King, 
arXiv:1503.03306 [hep-ph]. 

\bibitem{Chen:2003zv} M.~-C.~Chen and K.~T.~Mahanthappa, 
Int.\ J.\ Mod.\ Phys.\ A \textbf{18}, 5819 (2003) [hep-ph/0305088].

\bibitem{King:2013eh} S.~F.~King and C.~Luhn, 
Rept.\ Prog.\ Phys.\ \textbf{76}, 056201 (2013) [arXiv:1301.1340 [hep-ph]].

\bibitem{T7} C.~Luhn, S.~Nasri and P.~Ramond, 
Phys.\ Lett.\ B \textbf{652}, 27 (2007) [arXiv:0706.2341 [hep-ph]];
C.~Hagedorn, M.~A.~Schmidt and A.~Y.~Smirnov, 
Phys.\ Rev.\ D \textbf{79}, 036002 (2009) [arXiv:0811.2955 [hep-ph]];
Q.~H.~Cao, S.~Khalil, E.~Ma and H.~Okada, 
Phys.\ Rev.\ Lett.\ \textbf{106}, 131801 (2011) [arXiv:1009.5415 [hep-ph]]; 
C.~Luhn, K.~M.~Parattu and A.~Wingerter, 
JHEP \textbf{1212}, 096 (2012) [arXiv:1210.1197 [hep-ph]]; J.~Kile,
M.~J.~P\'erez, P.~Ramond and J.~Zhang, 
JHEP \textbf{1402}, 036 (2014) [JHEP \textbf{1411}, 158 (2014)]
[arXiv:1311.4553 [hep-ph]]; Y.~Kajiyama, H.~Okada and K.~Yagyu, 
JHEP \textbf{1310}, 196 (2013) [arXiv:1307.0480 [hep-ph]]; V.~V.~Vien and
H.~N.~Long, 
JHEP \textbf{1404}, 133 (2014) [arXiv:1402.1256 [hep-ph]]; C.~Bonilla,
S.~Morisi, E.~Peinado and J.~W.~F.~Valle, 
arXiv:1411.4883 [hep-ph]; J.~Kile, M.~J.~P\'erez, P.~Ramond and J.~Zhang, 
Phys.\ Rev.\ D \textbf{90}, no. 1, 013004 (2014) [arXiv:1403.6136 [hep-ph]];
A.~E.~C\'{a}rcamo Hern\'{a}ndez and R.~Martinez, arXiv:1501.07261 [hep-ph].


\bibitem{Perez:2007rm} P.~Fileviez Perez, 
Phys.\ Lett.\ B \textbf{654} (2007) 189 [hep-ph/0702287]. 

\bibitem{FileviezPerez:2007nh} P.~Fileviez P\`{e}rez, 
In *Karlsruhe 2007, SUSY 2007* 678-681 [arXiv:0710.1321 [hep-ph]]. 

\bibitem{Perez:2008ry} P.~Fileviez P\`{e}rez, H.~Iminniyaz and Germ\'{a}%
n~Rodrigo, 
Phys.\ Rev.\ D \textbf{78}, 015013 (2008) [arXiv:0803.4156 [hep-ph]]. 


\bibitem{Georgi:1974sy} H.~Georgi and S.~L.~Glashow, 
Phys.\ Rev.\ Lett.\ \textbf{32}, 438 (1974).

\bibitem{Georgi:1979df} H.~Georgi and C.~Jarlskog, 
Phys.\ Lett.\ B \textbf{86}, 297 (1979).

\bibitem{Frampton:1979} P.~Frampton, S.~Nandi and J.~Scanio, Phys.\ Lett.\ B 
\textbf{85}, 225 (1979). 

\bibitem{Ellis:1979} J.~R.~Ellis and M.~K.~Gaillard, 
Phys.\ Lett.\ B \textbf{88}, 315 (1979).

\bibitem{Nandi:1980sd} S.~Nandi and K.~Tanaka, 
Phys.\ Lett.\ B \textbf{92}, 107 (1980).

\bibitem{Frampton:1980} P.~H.~Frampton, 
Phys.\ Lett.\ B \textbf{89}, 352 (1980).

\bibitem{Langacker:1980js} P.~Langacker, 
Phys.\ Rept.\ \textbf{72}, 185 (1981). 

\bibitem{Kalyniak:1982pt} P.~Kalyniak and J.~N.~Ng, 
Phys.\ Rev.\ D \textbf{26}, 890 (1982). 

\bibitem{Giveon:1991} A.~Giveon, L.~J.~Hall and U.~Sarid, 
Phys.\ Lett.\ B \textbf{271}, 138 (1991).


\bibitem{Dorsner:2007fy} I.~Dorsner and I.~Mocioiu, 
Nucl.\ Phys.\ B \textbf{796}, 123 (2008) [arXiv:0708.3332 [hep-ph]]. 




\bibitem{Khalil:2013ixa} S.~Khalil and S.~Salem, 
Nucl.\ Phys.\ B \textbf{876}, 473 (2013) [arXiv:1304.3689 [hep-ph]]. 


\bibitem{Dorsner:2006dj} I.~Dorsner and P.~Fileviez Perez, 
Phys.\ Lett.\ B \textbf{642}, 248 (2006) [hep-ph/0606062].

\bibitem{Li:1973mq} L.~-F.~Li, 
Phys.\ Rev.\ D \textbf{9}, 1723 (1974). 


\bibitem{Grimus:2000vj} W.~Grimus and L.~Lavoura, 
JHEP \textbf{0011}, 042 (2000) [hep-ph/0008179]. 

\bibitem{Alvarado:2012xi} C.~Alvarado, R.~Martinez and F.~Ochoa, 
Phys.\ Rev.\ D \textbf{86}, 025027 (2012) [arXiv:1207.0014 [hep-ph]]. 


\bibitem{Hernandez:2013mcf} A.~E.~Carcamo Hernandez, R.~Martinez and
F.~Ochoa, 
Phys.\ Rev.\ D \textbf{87}, no. 7, 075009 (2013) [arXiv:1302.1757 [hep-ph]]. 


\bibitem{6HDMA4} A.~E.~C\'{a}rcamo Hern\'{a}ndez, I. d. M. Varzielas, S. G.
Kovalenko, H. P\"as and I. Schmidt, 
Phys.\ Rev.\ D \textbf{88}, 076014 (2013) [arXiv:1307.6499 [hep-ph]]

\bibitem{TM} 
W.~Grimus and L.~Lavoura, 
JHEP \textbf{0809}, 106 (2008) [arXiv:0809.0226 [hep-ph]]; 
W.~Grimus and L.~Lavoura, 
Phys.\ Lett.\ B \textbf{671}, 456 (2009) [arXiv:0810.4516 [hep-ph]]; 
C.~S.~Lam, 
arXiv:0907.2206 [hep-ph]; 
C.~H.~Albright and W.~Rodejohann, 
Eur.\ Phys.\ J.\ C \textbf{62}, 599 (2009) [arXiv:0812.0436 [hep-ph]]; 
C.~H.~Albright, A.~Dueck and W.~Rodejohann, 
Eur.\ Phys.\ J.\ C \textbf{70} (2010) 1099 [arXiv:1004.2798 [hep-ph]]; 
S.~Antusch, S.~F.~King, C.~Luhn and M.~Spinrath, 
Nucl.\ Phys.\ B \textbf{856}, 328 (2012) [arXiv:1108.4278 [hep-ph]]; 
C.~C.~Li and G.~J.~Ding, 
Nucl.\ Phys.\ B \textbf{881}, 206 (2014) [arXiv:1312.4401 [hep-ph]]. 

















\bibitem{Auger:2012ar} EXO Collaboration, M.~Auger \emph{et~al.}, \newblock %
Phys.Rev.Lett. \textbf{109}, 032505 (2012), arXiv:1205.5608. 

\bibitem{Abt:2004yk} GERDA Collaboration, I.~Abt \emph{et~al.}, \newblock %
(2004), arXiv:hep-ex/0404039. 

\bibitem{Ackermann:2012xja} GERDA Collaboration, K.-H. Ackermann \emph{et~al.%
}, \newblock (2012), Eur.\ Phys.\ J.\ C \textbf{73}, 2330 (2013)
[arXiv:1212.4067 [physics.ins-det]]. 

\bibitem{Alessandria:2011rc} F.~Alessandria \emph{et~al.}, \newblock (2011),
arXiv:1109.0494. 

\bibitem{KamLANDZen:2012aa} KamLAND-Zen Collaboration, A.~Gando \emph{et~al.}%
, \newblock Phys.Rev. \textbf{C85}, 045504 (2012), arXiv:1201.4664. 

\bibitem{Auty:2013:zz} EXO-200 Collaboration, D.~Auty, \newblock Recontres
de Moriond, http://moriond.in2p3.fr/ (2013).

\bibitem{Guiseppe:2011me} Majorana Collaboration, C.~Aalseth \emph{et~al.}, %
\newblock Nucl.Phys.Proc.Suppl. \textbf{217}, 44 (2011), arXiv:1101.0119. 

\bibitem{Reviewbetadecay} O.~Cremonesi and M.~Pavan, 
arXiv:1310.4692 [physics.ins-det]; W.~Rodejohann, 
J.\ Phys.\ G \textbf{39}, 124008 (2012) [arXiv:1206.2560 [hep-ph]];
A.~Barabash, \newblock(2012), arXiv:1209.4241; F.~F.~Deppisch, M.~Hirsch and
H.~P\"as, 
J.\ Phys.\ G \textbf{39}, 124007 (2012) [arXiv:1208.0727 [hep-ph]];
J.~Vergados, H.~Ejiri, and F.~Simkovic. 
Rept.Prog.Phys., \textbf{75}, 106301 (2012); A.~Giuliani and A.~Poves, 
Adv.\ High Energy Phys.\ \textbf{2012}, 1 (2012); 
S.~M.~Bilenky and C.~Giunti, 
arXiv:1411.4791 [hep-ph].

\bibitem{Bora:2012tx} K.~Bora, 
arXiv:1206.5909 [hep-ph]. 

\bibitem{Xing:2007fb} Z.~z.~Xing, H.~Zhang and S.~Zhou, 
Phys.\ Rev.\ D \textbf{77}, 113016 (2008) [arXiv:0712.1419 [hep-ph]]. 




\bibitem{Branco2010} G.~C.~Branco, D.~Emmanuel-Costa and C.~Simoes, 
Phys.\ Lett.\ B \textbf{690} (2010) 62 [arXiv:1001.5065 [hep-ph]];


\bibitem{CarcamoHernandez:2010im} A.~E.~C\'arcamo Hern\'andez and R.~Rahman, 
arXiv:1007.0447 [hep-ph]. 



\bibitem{Hernandez:2013hea} A.~E.~C\'arcamo Hern\'andez, R. Martinez and
F.~Ochoa, 
arXiv:1309.6567 [hep-ph]. 

\bibitem{King:2013hj} S.~F.~King, S.~Morisi, E.~Peinado and J.~W.~F.~Valle, 
Phys.\ Lett.\ B \textbf{724}, 68 (2013) [arXiv:1301.7065 [hep-ph]].

\bibitem{Hernandez:2014vta} A.~E.~C\'arcamo Hern\'andez, R. Martinez and
Jorge Nisperuza, 
Eur.\ Phys.\ J.\ C \textbf{75}, no. 2, 72 (2015) [arXiv:1401.0937 [hep-ph]].

\bibitem{Campos2014} M.~D.~Campos, A.~E.~C\'{a}rcamo Hern\'{a}ndez, H.~Pas
and E.~Schumacher, 
Phys.\ Rev.\ D \textbf{91}, no. 11, 116011 (2015) [arXiv:1408.1652
[hep-ph]]. 

\bibitem{Hernandez:2015dga} A.~E.~C\'{a}rcamo Hern\'{a}ndez,
I.~d.~M.~Varzielas and E.~Schumacher, 
arXiv:1509.02083 [hep-ph].

\bibitem{CarcamoHernandez2015} A.~E.~C\'{a}rcamo Hern\'{a}ndez, N.~Neill H
and I.~d.~M.~Varzielas, 
arXiv:1511.07420 [hep-ph].

\bibitem{KhalilDelta27} M.~Abbas and S.~Khalil, 
Phys.\ Rev.\ D \textbf{91}, no. 5, 053003 (2015) [arXiv:1406.6716 [hep-ph]].


\bibitem{Ishimori:2014jwa} H.~Ishimori and S.~F.~King, 
Phys.\ Lett.\ B \textbf{735}, 33 (2014) [arXiv:1403.4395 [hep-ph]].

\bibitem{Xing2010} Z.~-z.~Xing, D.~Yang and S.~Zhou, 
Phys.\ Lett.\ B \textbf{690}, 304 (2010) [arXiv:1004.4234 [hep-ph]].

\bibitem{Branco2012} 
G.~C.~Branco, H.~R.~C.~Ferreira, A.~G.~Hessler and J.~I.~Silva-Marcos, 
JHEP \textbf{1205}, 001 (2012) [arXiv:1101.5808 [hep-ph]].

\bibitem{Bhattacharyya:2012pi} G.~Bhattacharyya, I.~de Medeiros Varzielas
and P.~Leser, 
Phys.\ Rev.\ Lett.\ \textbf{109}, 241603 (2012) [arXiv:1210.0545 [hep-ph]]. 

\bibitem{CarcamoHernandez:2012xy} A.~E.~C\'{a}rcamo Hern\'{a}ndez,
C.~O.~Dib, N.~Neill H and A.~R.~Zerwekh, 
JHEP \textbf{1202}, 132 (2012) [arXiv:1201.0878 [hep-ph]].


\bibitem{Vien:2014ica} V.~V.~Vien and H.~N.~Long, 
arXiv:1408.4333 [hep-ph].

\bibitem{Ishimori:2014nxa} H.~Ishimori, S.~F.~King, H.~Okada and
M.~Tanimoto, 
Phys.\ Lett.\ B \textbf{743}, 172 (2015) [arXiv:1411.5845 [hep-ph]]. 

%
















\bibitem{Senjanovic:2009kr} G.~Senjanovic, 
AIP Conf.\ Proc.\ \textbf{1200}, 131 (2010) [arXiv:0912.5375 [hep-ph]]. 


\bibitem{Randall:1995sh} L.~Randall and C.~Csaki, 
In *Palaiseau 1995, SUSY 95* 99-109 [hep-ph/9508208]. 

\bibitem{DeRomeri:2011ie} V.~De Romeri, M.~Hirsch and M.~Malinsky, 
Phys.\ Rev.\ D \textbf{84}, 053012 (2011) [arXiv:1107.3412 [hep-ph]]. 

\bibitem{Arbelaez:2013hr} C.~Arbelaez, R.~M.~Fonseca, M.~Hirsch and
J.~C.~Romao, 
Phys.\ Rev.\ D \textbf{87}, no. 7, 075010 (2013) [arXiv:1301.6085 [hep-ph]]. 

\bibitem{Nath:2006ut} P.~Nath and P.~Fileviez Perez, 
Phys.\ Rept.\ \textbf{441}, 191 (2007) [hep-ph/0601023]. 

\bibitem{Abe:2013lua} K.~Abe \textit{et al.} [Super-Kamiokande
Collaboration], 
Phys.\ Rev.\ Lett.\ \textbf{113}, no. 12, 121802 (2014) [arXiv:1305.4391
[hep-ex]]. 


\bibitem{Biggio:2010me} C.~Biggio and L.~Calibbi, 
arXiv:1007.3750 [hep-ph]. 

\bibitem{LHC-color-8} G. Aad et al. (ATLAS Collaboration), Eur. Phys. J. C 
\textbf{73}, 2263 (2013); \textit{ibid.}, \textbf{71}, 1828 (2011).


\bibitem{Cirelli:2006} M. Cirelli, N. Fornengo, and A. Strumia, Nucl. Phys.
B \textbf{753}, 178 (2006); M. Cirelli, A. Strumia, and M. Tamburini, ibid.
B \textbf{787}, 152 (2007).


\bibitem{Khalil:2014gba} S.~Khalil, S.~Salem and M.~Allam, 
Phys.\ Rev.\ D \textbf{89}, no. 9, 095011 (2014) [arXiv:1401.1482 [hep-ph]]. 

\bibitem{Color-8-LHC-earlier} J.M. Arnold and B. Fornal, Phys. Rev. D 
\textbf{85}, 055020 (2012) ; S. I. Bityukov and N. V. Krasnikov, Mod. Phys.
Lett. A \textbf{12}, 2011 (1997); Y. Bai and B. A. Dobrescu, J. High Energy
Phys. \textbf{07}, 100 (2011); M. Gerbush, T. J. Khoo, D. J. Phalen, A.
Pierce, and D. Tucker-Smith, Phys. Rev. D \textbf{77}, 095003 (2008); X.-G.
He, G. Valencia, and H. Yokoya, J. High Energy Phys. \textbf{12}, 030
(2011); A. Idilbi, C. Kim, and T. Mehen, Phys. Rev. D \textbf{82}, 075017
(2010); C. Kim and T. Mehen, Phys. Rev. D \textbf{79}, 035011 (2009).

\bibitem{Capozzi:2013csa} F.~Capozzi, G.~L.~Fogli, E.~Lisi, A.~Marrone,
D.~Montanino and A.~Palazzo, 
Phys.\ Rev.\ D \textbf{89}, 093018 (2014) [arXiv:1312.2878 [hep-ph]].


\bibitem{Blanchet:2008cj} S.~Blanchet and P.~Fileviez Perez, 
JCAP \textbf{0808}, 037 (2008) [arXiv:0807.3740 [hep-ph]]. 

\bibitem{Discret-Group-Review} P. Ramond, \textit{Group Theory: A
Physicist's Survey}, Cambridge University Press, UK (2010); C.~Luhn,
S.~Nasri and P.~Ramond, 
J.\ Math.\ Phys.\ \textbf{48}, 123519 (2007) [arXiv:0709.1447 [hep-th]].
\end{thebibliography}
\end{document}